\documentclass [12pt]{article}
\usepackage {amssymb}
\usepackage {amsmath}
\usepackage [cp1251]{inputenc}
\usepackage [english]{babel}
\usepackage{graphicx}
\usepackage{cite}
\usepackage {longtable}

\title{Evolution of gas-filled pore in bounded particles.}

\author{ $^{1,2}$\textbf{V. V. Yanovsky},$^{1}$\textbf{M. I. Kopp},$^{1}$\textbf{M. A. Ratner}}

\begin{document}

\maketitle

$^{1}$ \textit{Institute for Single Cristals, National Academy of Science of Ukraine, Nauki Ave 60, 61001 Kharkiv, Ukraine}

$^{2}$\textit{Kharkiv National Karazin University, Svobody Sq. 4, 61000 Kharkiv, Ukraine}

 \abstract{In the present work, evolution of gas-filled pore inside spherical nanoshells is considered. On the supposition that diffusion fluxes are quasistationary, the nonlinear equation system is obtained analytically, that describes completely the behaviour of gas-filled pore and matrix shell. Two limiting cases are considered: the case when the pore is small as compared to the matrix shell and the case of the pore close to the matrix shell boundary. The characteristic regularities of pore behaviour are established.}

\section{Introduction}
One of the most important problems of contemporary material science is investigation of onset and development of gas porosity in materials. The creation of materials with improved radiation hardness is important for development of atomic energetics development as well as for other sectors of industry. Along with vacancy pores, gas-filled pores were discovered forming due to irradiating metals by quick neutron or charged particle fluxes in accelerators. For the first time, theoretical investigation of these problems was performed in the works \cite{1s}-\cite{7s}. In the same works, the growth of pores filled with noble gases was considered as applied to material swelling, that is, to a large degree, connected with pore coalescence. The physical cause of material swelling as a consequence of gas porosity consists in absorbing of thermal vacancies at redistribution of pores during the coalescence. Pore behaviour becomes even more complicated if it is filled chemically active gas (or gases)  that at coalescence temperatures can interact matrix material or other gases, forming inside the pore one or several gaseous compounds). Such situation can take place, for example, under irradiation. At that, fragments in the form of chemically active gas molecules are formed in the material. The process of gas-filled bubble formation can, probably, occur in many materials, since practically all real materials contain interstitial impurities in the form of oxide, carbide, nitride, and other phases \cite{4s}.

Another up-to-date trend connected with investigation of gas porosity relates to the creation of new nano- and mesomaterials.  Such materials are formed via consolidation of nano-and mesoscale particles that, initially, possess complex defect structure. Properties of such particles to a large degree are determined just by this defect structure \cite{8s}-\cite{17s}. Regularities of diffusion growth, healing and motion of such defects in nanoparticles present an important problem for further compactification of nanoparticles and creation of new materials.

Such materials find important applications in optical spectroscopy, biomedicine, electronics and other areas \cite{18s}-\cite{19s}.
Creation of the theory of diffusive interaction of pores in bounded media, for example, in spherical particles, is an exceptionally complicated task.

In bounded particles of the matrix, the influence of close boundary complicates strongly pore behaviour.  Closeness of boundary leads to principally different pore behaviour as compared to that in unbounded materials. It is worse to note, that pore formation in spherical nanoshells was discovered relatively recently \cite{8s}. In the review \cite{20s} the results are presented of theoretical and numerical investigations related to formation and disappearing of pores in spherical and cylindrical nanoparticles. Great attention in \cite{20s} is paid to the problem of hole nanoshell stability, i.e. to the case when in the nanoparticle center large vacy pores are situated.

Analytical theory of diffusive interaction of the nanoshell and the pore situated at arbitrary distance from particle center was considered in the work  \cite{21s}. Here, the behaviour of vacancy pore inside solid matrix of spherical shape. With the supposition of quasiequilibrium of diffusive fluxes, the equations have been obtained analytically for the change of the radii of pore and spherical granule as well as of center-to-center distance between the pore and the granule. The absence of critical pore size has been demonstrated unlike the case of inorganic matrix. In general case, pore in such particles dissolves diffusively, while diminishing in size and shifting towards granule center.

In the present work, a simple case in considered of zero diffusion coefficient of the gas in the matrix. It has been shown that the behaviour of gas-filled pore is qualitatively different from that of vacancy pore in spherical matrix. Thus, unlike vacancy pore, the gas-filled one is of stable size, that is determined by the gas density.  Asymptotic regimes as well as main regularities of the gas-filled pore behaviour has been established.

\section{Evolution equations of gas-filled pore}

\begin{figure}
  \centering
  \includegraphics[width=7 cm]{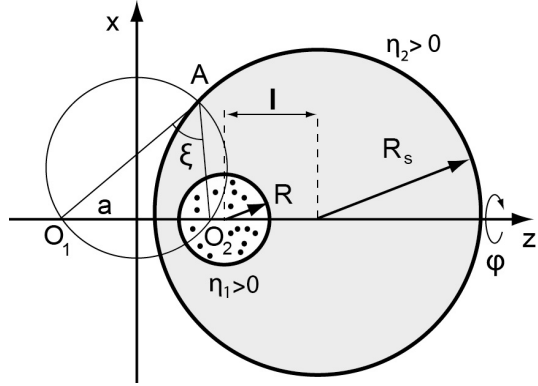}\\
  \caption{Gas-filled pore in spherical granule in bispherical coordinate system. Pore and granule surfaces in this system are coordinate planes $\eta =\textrm{const}.$}  \label{fg1}
\end{figure}

Let us consider the spherical granule of the radius $R_s$ containing the gas-filled pore of the radius $R < R_{s}$ (see Fig.\ref{fg1}).
Let us designate the initial values of these radii (at time moment $t=0$)  as $R_s(0)$ and $R(0)$ correspondingly.
Suppose that granule and pore centers are separated from each other by the distance $l$. We assume that the gas can be found only inside the pore and neglect gas diffusion through pore boundaries. We are interested in pore and granule evolution under the influence of diffusive vacancy fluxes. The complete description of such evolution assumes the knowledge of pore and granule size change with time as well as of time change of their center-to-center distances. In order to obtain the equations describing such evolution, the boundary conditions are required, that are determined by equilibrium vacancy concentrations near the pore and granule surfaces. The equilibrium vacancy concentration near a spherical pore surface is determined, with the account of gas pressure, by the relation (see e.g. \cite{7s},\cite{22s} ),
\begin{equation}\label{eq1}
    c_{R}=c_{V}\exp\left(\frac{2\gamma\omega}{kTR}-\frac{P\omega}{kT}\right)\,,
\end{equation}
where $c_V$ is equilibrium vacancy concentration near the plane surface, $\gamma$  is the surface energy per unit area, $T$ is granule temperature, $\omega$ is the volume per lattice site  $\omega$ is the atomic volume of a vacancy, $P$ is gas pressure inside the pore satisfying the equation of state of ideal gas:
\[P\cdot \frac{4\pi}{3}\cdot R^3=N_g kT,\]
here $N_g$ is the gas quantity inside the pore.
In the same way, the equilibrium concentration of vacancies near the spherical granule free surface is determined. At this, taking into account that of interest are small granules of nano- and meso- sizes, it is natural to assume the smallness of the external pressure as compared to the Laplass pressure.  Then the equilibrium vacancy concentration near the spherical granule free surface is determined as
\begin{equation}\label{eq2}
   c_{R_{s}}=c_{V}\exp\left(-\frac{2\gamma\omega}{kTR_s}\right)\,,
\end{equation}
These concentration values will determine vacancy fluxes. In the further consideration,
we will suppose that equilibrium concentrations adjust quickly to the change of pore
and granule sizes. In other words, equilibrium concentrations tune themselves to pore
and granule size change. Certainly, the problem remains extremely complicated. For
the sake of simplicity, it is natural to make one more assumption, namely, to suppose
that stationary fluxes of vacancies inside granule are quickly established. There are
two arguments in favour of this.
First of all, even if one gets out of the limits of such assumption, vacancy distribution inside granule is unknown. Besides, in a number of cases, stationary fluxes are established quickly enough. The evaluation of characteristic time during which stationary fluxes are established gives $\tau \ll l^2/D$.
Under such assumptions, diffusion flux of vacancies onto pore and granule boundaries is determined by stationary diffusion equation and corresponding boundary conditions
\begin{equation}\label{eq3}
  \Delta c =0 ,
\end{equation}
\[c(r)|_{r=R}=c_{R} ,\]
\[c(r)|_{r=R_{s}}=c_{R_{s}} .\]
The geometry of pore and granule boundaries dictates the use of bispherical coordinate system  \cite{23s}, as the most convenient one. In bispherical coordinate system  (see Fig.\ref{fg1}) each point  $A$ of the space is matched to three numbers $(\eta,\xi,\varphi)$, where $\eta=\ln(\frac{|AO_1|}{|AO_2|})$, $\xi=\angle O_1AO_2$, $\varphi$ is polar angle.
  Let us cite relations connecting bi-spherical coordinates with Cartesian ones:
\begin{equation}\label{eq4}
  x=\frac{a\cdot \sin\xi\cdot\cos\varphi}{\cosh\eta-\cos\xi}, \quad
  y=\frac{a\cdot \sin\xi\cdot\sin\varphi}{\cosh\eta-\cos\xi}, \quad
  z=\frac{a\cdot \sinh\eta}{\cosh\eta-\cos\xi},
\end{equation}
where $a$ is the parameter, that at fixed values of pore and granule radii as well as of their center-to-center distance is determined by the relation
 $$a=\frac{\sqrt{[(l-R)^2-R_s^2][(l+R)^2-R_s^2]}}{2\cdot l}\,.$$
Pore and granule surfaces in such coordinate system are given by relations
\begin{equation}\label{eq5}
    \eta_1=\textrm{arsinh} \left(\frac{a}{R}\right), \quad
\eta_2=\textrm{arsinh} \left(\frac{a}{R_s}\right).
\end{equation}

These relations determine values of  $\eta_1$ and $\eta_2$ from pore and granule radii, while  $a$ includes additionally center-to center distance between the pore and the granule. In the bispherical coordinate system the equation determining vacancy concentration and boundary condition takes on a following form:
\begin{equation}\label{eq6}
\frac{\partial}{\partial\eta}\left(\frac{1}{\cosh\eta-\cos\xi}\frac{\partial c}{\partial\eta}\right)+
  \frac{1}{\sin\xi}\frac{\partial}{\partial\xi}\left(\frac{\sin\xi}{\cosh\eta-
  \cos\xi}\frac{\partial c}{\partial\xi}\right)+\frac{1}{(\cosh\eta-
  \cos\xi)\cdot \sin^2\xi}\frac{\partial^2
  c}{\partial\varphi^2}= 0
\end{equation}
\[c(\eta , \xi, \varphi)|_{\eta_1}=c_{R}\]
\[c(\eta , \xi, \varphi)|_{\eta_2}=c_{R_s}\]
Due to symmetry of the problem, vacancy concentration does not depend on variable $\varphi$. Consequently, equation (\ref{eq6}) is reduced to
\begin{equation}\label{eq7}
    \frac{\partial}{\partial\eta}\left(\frac{1}{\cosh\eta-\cos\xi}\frac{\partial c}{\partial\eta}\right)+
  \frac{1}{\sin\xi}\frac{\partial}{\partial\xi}\left(\frac{\sin\xi}{\cosh\eta-
  \cos\xi}\frac{\partial c}{\partial\xi}\right)= 0
\end{equation}
Let us perform substitution for the required function $c(\eta,\xi)=\sqrt{\cosh\eta-\cos\xi}\cdot
F(\eta,\xi)$ gives us equation for function  $F(\eta,\xi)$ in the following form:
\begin{equation}\label{eq8}
 \frac{\partial^2F}{\partial\eta^2}+\frac{1}{\sin\xi}\frac{\partial}{\partial\xi}
 \left(\sin\xi\frac{\partial F}{\partial\xi}\right)-\frac14F=0\,,
\end{equation}
 Let us try solution of the equation by the method of separation of variables:  $F(\eta,\xi)= F_1(\eta)\cdot F_2(\xi)$. As a result, the following equations are obtained:
\[\frac{d^2F_1}{d\eta^2}=\left(k+\frac{1}{2} \right)^2\cdot F_1 ,\]
\[ \frac{1}{\sin\xi}\frac{d}{d\xi}
 \left(\sin\xi\frac{d F_2}{d\xi}\right)=-k\cdot(k+1)\cdot F_2\,.\]
Here parameter $k$ is a separation constant.
The solution of these equations can be easily found, taking into account that the second one coincides with Legendre equation. Then, general solution can be written down in the form:
\begin{equation}\label{eq9}
c(\eta,\xi) = \sqrt{\cosh\eta-\cos\xi}\times$$
$$\times\sum_{k=1}^{\infty} (A_k\cdot\exp(k+1/2\eta)\cdot
P_k(\cos(\xi))+B_k\cdot\exp(-(k+1/2\eta))\cdot P_k(\cos(\xi)))\,,
\end{equation}
where $A_k$ and $B_k$ are, arbitrary constants, and $P_k(x)$  are Legendre polynoms.
\[P_k(x) = \frac1{2^k\cdot k!}\frac{d^k}{dx^k}(x^2-1)^k,\,\quad P_0(x)\equiv 1.\]
We still have to determine the values of arbitrary constants from boundary conditions and find boundary problem solution  (\ref{eq6}) as
\begin{equation}\label{eq10}
c(\eta, \xi ) =
\sqrt{2(\cosh\eta-\cos\xi)}\left\{{c_R}\sum_{k=0}^\infty
\frac{\sinh(k+1/2)(\eta-\eta_2)}{\sinh(k+1/2)(\eta_1-\eta_2)}\exp(-(k+1/2)\eta_1)P_k(\cos\xi)-
\right.$$
 $$ \left.- c_{R_s}\sum_{k=0}^\infty
\frac{\sinh(k+1/2)(\eta-\eta_1)}{\sinh(k+1/2)(\eta_1-\eta_2)}\exp(-(k+1/2)\eta_2)P_k(\cos\xi)
\right\}\,.
\end{equation}
Let us note, that here boundary concentration $c_R$ is expressed through $\eta_1$ and $a$, and $c_{R_s}$ through $\eta_2$ and $a$. This solution determines stationary vacancy concentration anywhere inside spherical granule of  radius  $R_s$ and outside pore of radius $R$. However, the knowledge of vacancy concentration allows one to find vacancy fluxes onto the pore as well as onto granule boundary at the given positions of granule and pore. These fluxes cause change size and position of pore. With account of this, one can write down the equations for the time change of pore and granule radii as well as of their center-to-center distance. Vacancy flux is determined by the first Fick's low as
\begin{equation}\label{eq11}
\vec{j}=-\frac {D}{\omega} \nabla c\,,
\end{equation}
where $D$ is diffusion coefficient. Let denote the outer pore surface normal as  $\vec{n}$. Then vacancy flux onto pore surface is determined by scalar product  $\vec{n} \cdot \vec{j}|_{\eta=\eta_1}$.  Let us write down the expression for vacancy flux onto unit area of pore surface using the expression for gradient in bispherical coordinates \cite{23s}
\begin{equation}\label{eq12}
\vec{n} \cdot \vec{j}|_{\eta=\eta_1}=\frac {D}{\omega} \cdot
\frac{\cosh\eta_1-\cos\xi}{a}\frac{\partial
c}{\partial\eta}\left|_{\eta=\eta_1}\right.\,.
\end{equation}
Similar expression determines vacancy flux onto unit area of granule surface
\begin{equation}\label{eq13}
\vec{n} \cdot \vec{j}|_{\eta=\eta_2}=\frac {D}{\omega} \cdot
\frac{\cosh\eta_2-\cos\xi}{a}\frac{\partial
c}{\partial\eta}\left|_{\eta=\eta_2}\right.\,.
\end{equation}
Here $\vec{n}$ is granule surface normal. Evidently, the total vacancy flux onto pore surface determines the rate of pore volume change. It is natural to suppose, that surface diffusion, whose diffusion coefficient usually much exceeds that of the bulk, is in time to restore spherical shape of the pore and the granule. Thus, it is easy to write down the equation for pore volume change in the form
\[\dot{R}=-\frac{\omega}{4\pi R^2} \oint\vec{n}\vec{j}|_{\eta=\eta_1} \, dS\]
In the same way one obtains the equation that determines granule radius:
\[\dot{R_s}=-\frac{\omega}{4\pi R_s^2}\oint \vec{n}\vec{j}|_{\eta=\eta_2} dS\]
After substitution of the exact solution and performing integration, one obtains an equation for pore radius change with time:
\begin{equation}\label{eq14}
    \dot{R}=-\frac{D}{R}\left[\frac{c_R}{2}
+\sinh\eta_1\cdot(c_R\cdot(\Phi_1+\Phi_2)-2c_{R_s}\cdot\Phi_2)\right],
\end{equation}
where functions $\Phi_1$ and $\Phi_2$ are introduced, that consist of the sum of exponential series:

\[\Phi_1=\sum_{k=0}^\infty \frac{ e^{-(2k+1)\eta_1}}{e^{(2k+1)(\eta_1-\eta_2)}-1}, \quad \Phi_2=\sum_{k=0}^\infty \frac{ e^{-(2k+1)\eta_2}}{e^{(2k+1)(\eta_1-\eta_2)}-1}\]

The details of the derivation are given in the appendix. Here $\eta_1$ and $\eta_2$ are expressed through pore and granule radii in correspondence with relations (\ref{eq5}), while $c_R$ and $c_{R_s}$ through (\ref{eq1}), (\ref{eq2}). Thus, the right part of this equation depends nonlinearly on $R$, ${R_s}$ and $l$. In a similar way we obtain equation
\begin{equation}\label{eq15}
    \dot{R_s}=-\frac{D}{R_s}\left[\frac{c_{R_s}}{2}
+\sinh\eta_2\cdot(2c_{R}\cdot\Phi_2 - c_{R_s}\cdot(\Phi_2+\Phi_3)) \right]\,,
\end{equation}
where the following definition for the function $\Phi_3$ is introduced:
\[\Phi_3=\sum_{k=0}^\infty \frac{ e^{-(2k+1)(2\eta_2-\eta_1)}}{e^{(2k+1)(\eta_1-\eta_2)}-1}=\sum_{k=0}^\infty \frac{ e^{-(2k+1)\eta_3}}{e^{(2k+1)(\eta_1-\eta_2)}-1}\]

In order to obtain a closed set of equations determining granule and pore evolution, one needs to complement these equations with one for the rate of changing center-to-center distance between the pore and the granule.  Of course, the displacement rate of vacancy pore relative to granule center is also determined by diffusion fluxes of vacancies onto pore surface (see e.g. \cite{7s},\cite{22s} ). In the present case, the displacement rate is determined by relation
\begin{equation}\label{eq16}
\vec{v}=-\frac{3\omega}{4\pi R^2}\oint \vec{n}(\vec{n}\cdot\vec{j}_v)|_{\eta=\eta_1} dS.
\end{equation}
 Using again the exact solution (\ref{eq10}) and performing integration (see Appendix), one obtains:
\begin{equation}\label{eq17}
\vec{v}=\vec{e_z}\cdot\frac{3 D}{R}\times$$
$$\times
\left[\sinh^2\eta_1 \cdot(c_R\cdot
(\widetilde{\Phi}_1+\widetilde{\Phi}_2)-2c_{R_s}\cdot \widetilde{\Phi}_2)-\frac{1}{2}\sinh2\eta_1\cdot(c_R\cdot(\Phi_1+\Phi_2)-2c_{R_s}\cdot \Phi_2)\right]
\end{equation}
Here new functions $\widetilde{\Phi}_1$ and $\widetilde{\Phi}_2$ are defined:

\[\widetilde{\Phi}_1=\sum_{k=0}^\infty \frac{(2k+1)e^{-(2k+1)\eta_1}}{e^{(2k+1)(\eta_1-\eta_2)}-1}, \quad \widetilde{\Phi}_2=\sum_{k=0}^\infty \frac{(2k+1)e^{-(2k+1)\eta_2}}{e^{(2k+1)(\eta_1-\eta_2)}-1}\,.\]

Taking into account that displacement rate along  $z$ coincides with $dl/dt$, let us write down the equation in the final form
\begin{equation}\label{eq18}
  \frac{dl}{dt} =  \frac{3 D}{R}\times$$
$$\times
\left[\sinh^2\eta_1 \cdot(c_R\cdot
(\widetilde{\Phi}_1+\widetilde{\Phi}_2)-2c_{R_s}\cdot \widetilde{\Phi}_2)-\frac{1}{2}\sinh2\eta_1\cdot(c_R\cdot(\Phi_1+\Phi_2)-2c_{R_s}\cdot \Phi_2)\right]
\end{equation}
The obtained equation set  (\ref{eq14}), (\ref{eq15}) and (\ref{eq18}) determines completely evolution of the gas-filled pore and the granule with time. In the limiting case when the gas is absent, $P=0$ (vacancy pore), equations (\ref{eq14}), (\ref{eq15}) and (\ref{eq18}) agree with results of the work \cite{23s}. Let us discuss several general properties of the obtained equation set. First of all, it is clear that the volume of granule material does not change with time. Vacancies only carry away 'emptiness'. It is easy to establish this conservation law from the obtained equation set. It can be shown easily that
\[R_s(t)^2\dot{R}_s(t)-R(t)^2\dot{R}(t)=0\]
The validity of such conservation law is connected closely with current quasi stationary approximation. Vacancy  fluxes, that come out from the pore and from the granule are balanced with each other. Thus, the volumes of the pore and of the granule are connected with each other by an easy relation
\begin{equation}\label{eq19}
  R_s(t)^3 ={V+R(t)^3}
\end{equation}
where $V=R_s(0)^3 -R(0)^3$ is initial volume of granule material (multiplier  $4\pi /3$ is omitted for convenience).  From the very statement of the problem, the second conservation low follows, i.e. the low of conservation of gas amount inside the pore:  $m_g=\textrm{const}$ or $N_g=\textrm{const}$. The existence of conservation low (\ref{eq19}) enables us to reduce the number of unknown quantities. As a result, we obtain Cauchy problem for the system of two differential equations $R$ and $l$, whose solution describes evolution of gas-filled pore inside nanoparticle:
\begin{equation}\label{eq20}
\begin{cases}
\frac{d l}{d t}=\frac{3Dc_V}{R}\cdot\exp\left(\frac{2\gamma\omega}{kTR}-\frac{3\omega N_g}{4\pi R^3}\right) \cdot \left[\frac{a^2}{R^2} \cdot (\widetilde \Phi _1  + \widetilde \Phi _2 ) - \frac{a}{R} \cdot \sqrt {1 + \frac{a^2}{R^2}}  \cdot (\Phi _1  + \Phi _2 )\right]-\\
-\frac{6Dc_V}{R}\cdot\exp\left(-\frac{2\gamma\omega}{kTR_{s}}\right)\cdot \left[\frac{a^2}{R^2} \cdot \widetilde \Phi _2  - \frac{a}{R} \cdot \sqrt {1 + \frac{a^2}{R^2}}  \cdot \Phi _2 \right],\\
\frac{d R}{d t}=-\frac{Dc_V}{R}\cdot\exp\left(\frac{2\gamma\omega}{kTR}-\frac{3\omega N_g}{4\pi R^3}\right) \cdot \left[\frac{1}{2} + \frac{a}{R} \cdot (\Phi _1  + \Phi _2 )\right]+\\
+\frac{2Dc_V}{R}\cdot\exp\left(-\frac{2\gamma\omega}{kTR_{s}}\right)\cdot \frac{a}{R} \cdot \Phi _2,\\
  R_s =\sqrt[3]{V+R^3},\\
  R|_{t=0}=R(0),\\
  l|_{t=0}=l(0).
\end{cases}
\end{equation}
For the sake of convenience, let us make equation set (\ref{eq20}) dimensionless with characteristic length $R_0 = R(0)$ (that is pore radius at the initial time moment $t = 0$) and characteristic time $t_D = R_0^2/Dc_V$. Let us now go over to the following dimensionless variables:
\[r = \frac{R}{R_0},\quad r_s = \frac{R_{s}}{R_0},\quad L = \frac{l}{R_0},\quad {\tau} = \frac{t}{{t_D }},  \quad \alpha = \frac{a}{R_0},\quad  \frac{{2\gamma \omega }}{{kTR}} = \frac{A}{{r}}, \]
\[\frac{{2\gamma \omega }}{{kTR_s}} = \frac{A}{{r_s}},\quad A = \frac{{2\gamma \omega }}{{kTR_0}},\quad \frac{3\omega N_g}{4\pi R^3} = \frac{B}{r^3}, \quad B=\frac{3\omega N_g}{4\pi R_0^3}.\]

Ultimately, the equation system  (\ref{eq20}) can be rewritten in dimensionless form:

\begin{equation}\label{eq21}
\begin{cases}
  \frac{d L}{d \tau}=\frac{3\exp\left(\frac{A}{r}-\frac{B}{r^3}\right)}{r} \cdot \left[\frac{\alpha^2}{r^2} \cdot (\widetilde \Phi _1  + \widetilde \Phi _2 ) - \frac{\alpha}{r} \cdot \sqrt {1 + \frac{\alpha^2}{r^2}}  \cdot (\Phi _1  + \Phi _2 )\right]-\\
-\frac{6\exp\left(-\frac{A}{r_s}\right)}{r}\cdot \left[\frac{\alpha^2}{r^2} \cdot \widetilde \Phi _2  - \frac{\alpha}{r} \cdot \sqrt {1 + \frac{\alpha^2}{r^2}}  \cdot \Phi _2 \right],\\
\frac{d r}{d \tau}=-\frac{\exp\left(\frac{A}{r}-\frac{B}{r^3}\right)}{r} \cdot \left[\frac{1}{2} + \frac{\alpha}{r} \cdot (\Phi _1  + \Phi _2 )\right]+\frac{2\exp\left(-\frac{A}{r_s}\right)}{r}\cdot \frac{\alpha}{r} \cdot \Phi _2,\\
r_s=\sqrt[3]{V+r^3},\\
  r|_{\tau =0}=1,\\
  L|_{\tau =0}=\frac{l(0)}{R(0)}.
\end{cases}
\end{equation}
The obtained non-linear system of evolution equations (\ref{eq21})
\begin{figure}
  \centering
  \includegraphics[width=7 cm, height=7 cm]{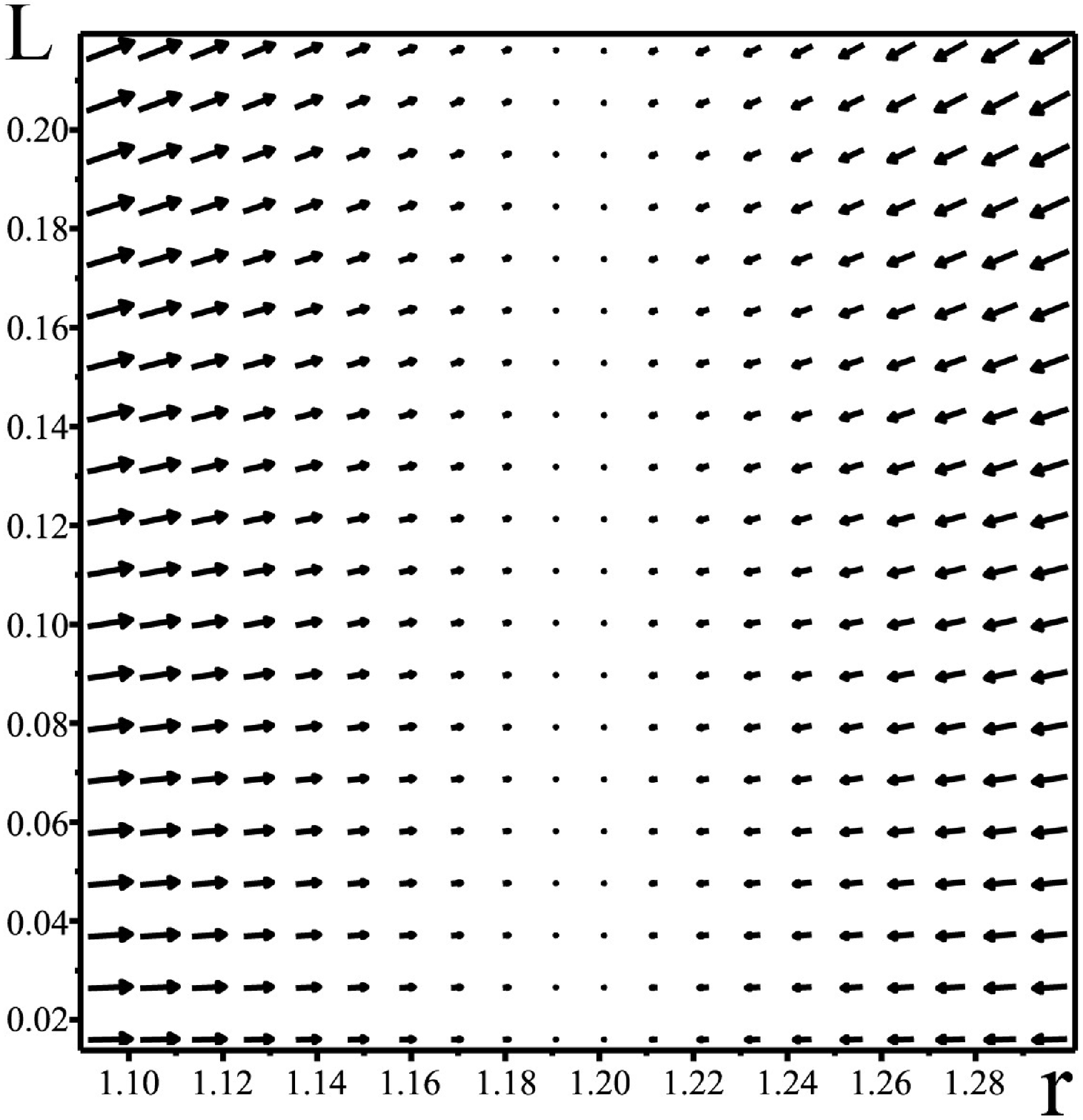}
	\includegraphics[width=7 cm, height=7 cm]{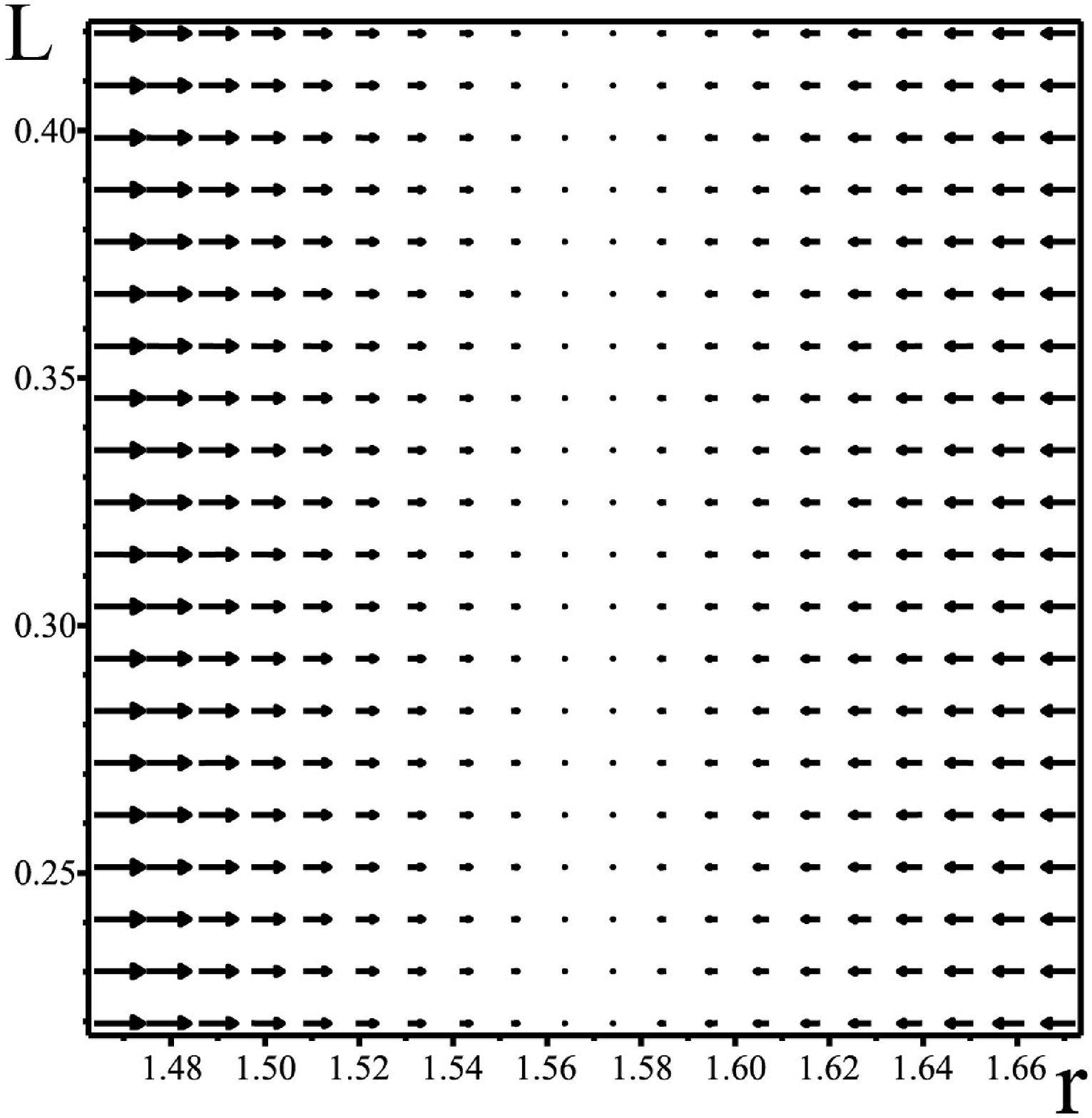}\\
	\caption{Phase portraits of trajectories in the plain $(r,L)$ are presented, obtained via numerical solution of Eqs. (\ref{eq21}). On the left: phase portrait for trajectories of "small" pores, obtained at initial conditions $r|_{t=0}=1$, $r_s|_{t=0}=100$,  $A=10^{-1}$ and  $B=0.25$; on the right: phase portrait for trajectories of "large" pores, obtained at initial conditions $r|_{t=0}=1$, $r_s|_{t=0}=1.5$,  $A=10^{-1}$ and  $B=0.25$ }
\label{fg2}
\end{figure}
is rather complicated. However we can analyse gas-filled pore evolution numerically via building vector field that mined by the right parts of equation system (\ref{eq21}). Corresponding vector field in the plain $(r,L)$ is demonstrated in Fig.\ref{fg2}. Here $r$ and $L$ are pore radius and position relative to the granule center correspondingly. Integral lines of this vector field determine phase portrait of equation system (\ref{eq21}).

The vector field for the case of "small" pores is shown at Fig. \ref{fg2} on the left, and for the "large" pores, on the right. The exact pore classification into "small" and "large" will be described below. It is easy to note that there is a limiting pore size $r_{cr}$, which a pore tries to accommodate during evolution. The size depends on the pore position either slightly or not at all. It can be understood from the physical point of view,if one takes into account that boundary conditions (\ref{eq1}), (\ref{eq2}) in this approximation do not depend on pore position. Pore motion, that is caused by vacancy fluxes onto the boundary of spherical surface, is limited by the value of gas pressure. Thus, after reaching the size $r_{cr} \approx  \sqrt{3 N_g kT/8 \pi \gamma}$, at which boundary conditions  become level due to gas pressure, the pore ceases changing its size and move.

Thus evolution of gas-filled pore consists in its tendency to reach some stationary size, while its position changes slowly and insignificantly. As pore size becomes close to its stationary value, pore motion is ceasing. For large pores, the direction of their motion depends on pore size. If the pore is larger then its stationary size, then, in the process of diminishing down to the stationary size, it moves towards the  granule center (see the right part of Fig. \ref{fg2}). If the pore is smaller  then its stationary size, then, in the process of growing up to the stationary size, it moves away from the  granule center (see the right part of Fig. \ref{fg2}).

\section{Asymptotic evolution modes}

Let us consider asymptotic modes of equation set (\ref{eq21}). Possible modes are determined by three dimensionless values: $R/R_s$, $l/R_s$  and  $R/l$. Let us suppose that the pore is situated at the distance $l$ from the granule center and its radius equals to $R$. The condition that such a pore is situated inside the granule leads to the purely geometrical inequality
\begin{equation}\label{eq22}
    R/R_s+l/R_s < 1
\end{equation}
Such inequality is held for all evolution modes of a gas-filled pore. In different cases, the mentioned above characteristic dimensionless values are of different order of smallness. Thus, the value $\delta=R/R_s < 1$   is always smaller the unity. The same relates also to $l/R_s < 1$. Assuming smallness of some values with account of the geometrical restriction, we obtain possible asymptotic modes. Below, we will discuss in more details asymptotic modes that can be realized. Besides, it is clear, that the character of pore evolution is influenced by  concentration ( or pressure) of the gas inside the pore. Here, three cases can be distinguished. The first is the case of high gas concentration, that leads to pore "swelling" up to the stationary size. Second case corresponds to such a value of gas concentration inside the pore, that initial radius and position of the pore do not change during evolution time. Third case relates to small gas concentration at which  decrease of the pore size down to some stationary value occurs.

\subsection{Small pores}

First of all, let us consider the case of small pores $R/R_s \ll 1$. At this, distance from the pore to granule center can take on different values.
Thus, the case is possible when
\[ R/R_s \ll 1, \quad l/R_s \ll 1,\].
Here, the relation between this values can vary. The possibility exists that
\[R/R_s \ll l/R_s  \Rightarrow R/l \ll 1\]
This means, that the distance from a small pore to the granule center is large as compared to granule radius. Thus, the mode exists when
\begin{equation}\label{eq23}
 1) \quad R/R_s \ll 1, \quad l/R_s \ll 1,\quad R/l \ll 1
\end{equation}

Of course, another disposition is possible, when a small pore is situated close to the granule center. In this case, the relation between the values is opposite:
\[R/R_s \gg l/R_s  \Rightarrow R/l \gg 1\]
Then, the next possible mode is determined by the relations of values
\begin{equation}\label{eq24}
   2) \quad R/R_s \ll 1, \quad l/R_s \ll 1,\quad R/l \gg 1
\end{equation}

\begin{figure}
  \centering
  \includegraphics[ height=7 cm]{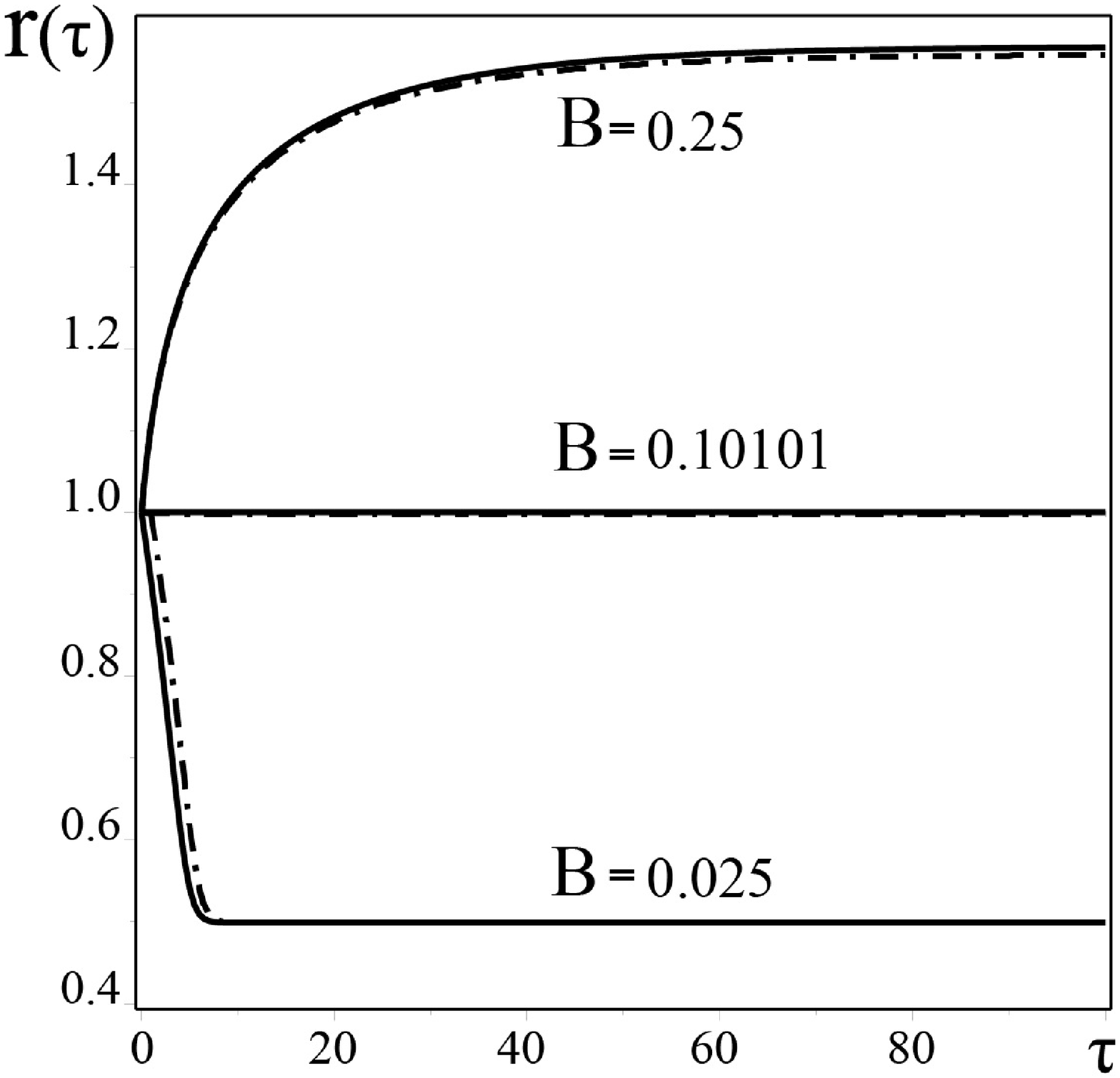}
	\includegraphics[ height=7 cm]{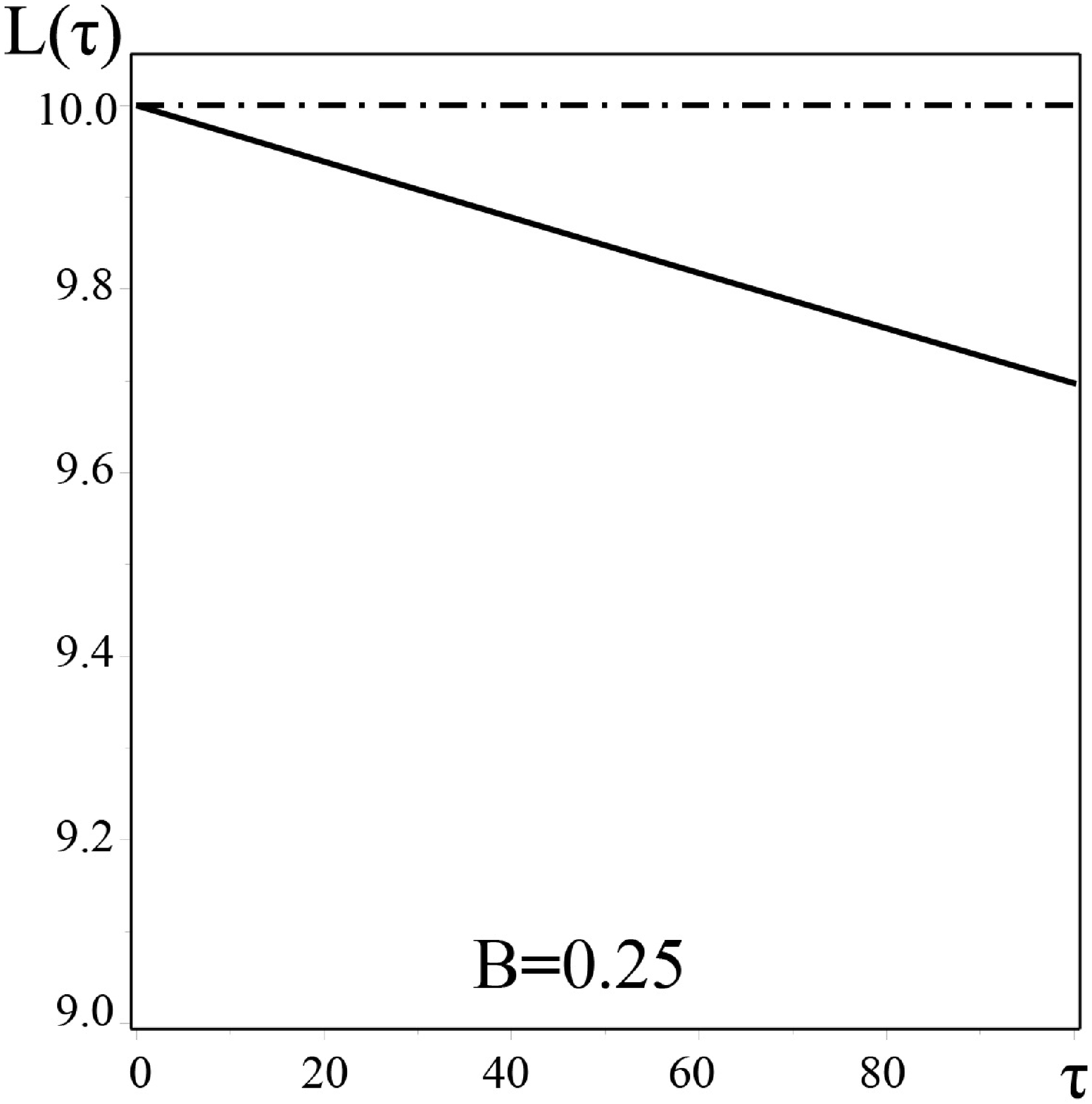}\\
	\includegraphics[ height=7 cm]{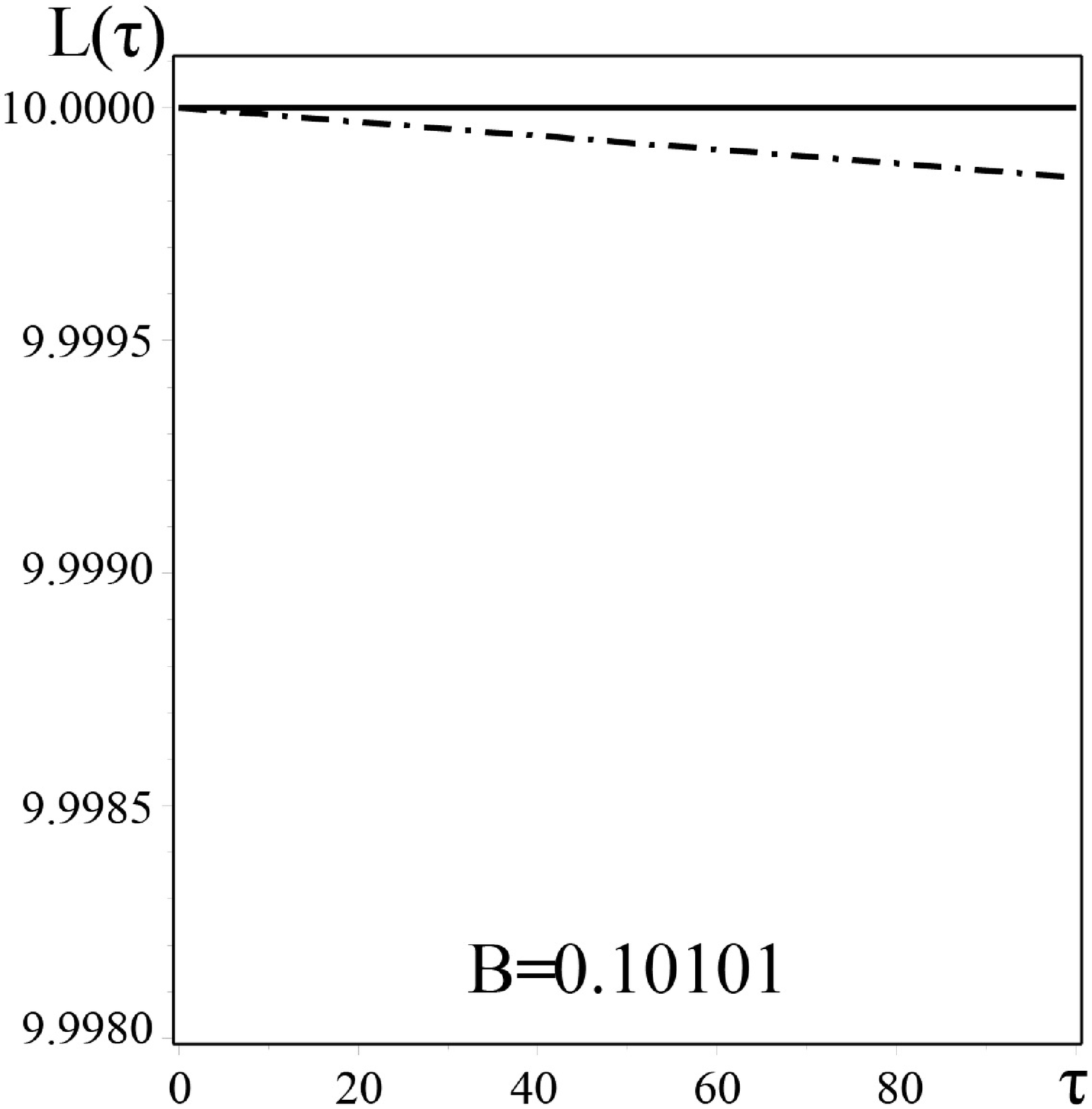}
	\includegraphics[ height=7 cm]{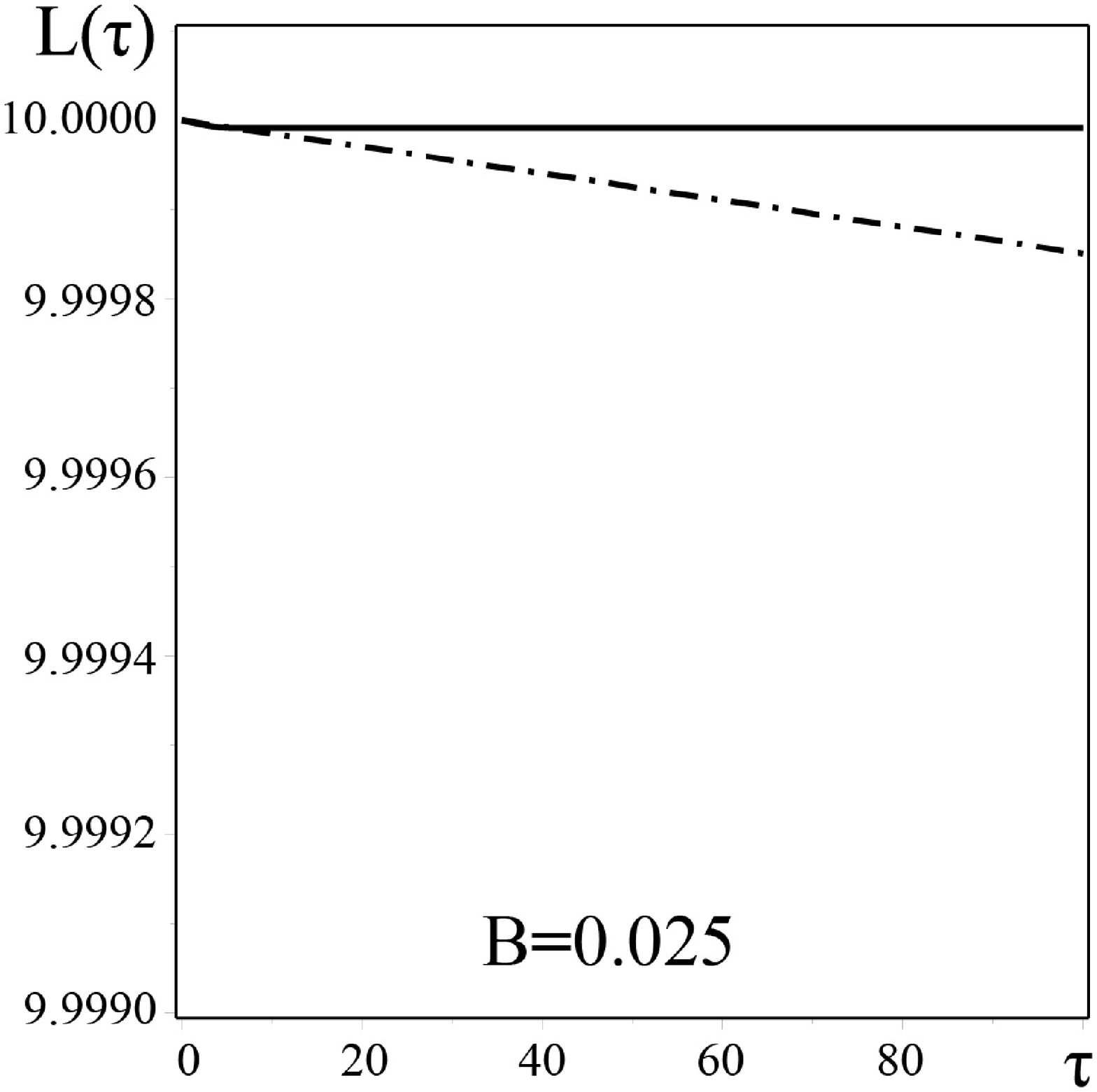}\\
 \caption{On the upper left, the dependence is shown of  pore radius $r$ on time $\tau$ for different values of gas parameter $B$: $B=0.25$, $B=0.10101$  and $B=0.025$. Solid line corresponds to the numerical solution of complete equation set (\ref{eq26}), dash-and-dot line corresponds the numerical solution of approximate equations (\ref{eq30})-(\ref{eq31});  on the upper right, the dependence is presented of distance $L$ on time $\tau$ for parameter $B=0.25$; below on the left, the dependence is shown of distance $L$ on time $\tau$ for parameter $B=0.10101$; below on the right, the dependence is shown of distance $L$ on time $\tau$ for parameter $B=0.025$. All solutions are obtained at initial conditions  $r|_{\tau=0}=1$, $r_s|_{\tau=0}=100$, $L|_{\tau=0}=10$ and $A=10^{-1}$.}
\label{fg3}
\end{figure}

Moreover, small pores can be situated at significant distance from the granule center that is comparable with granule size.  In this case, the next relations are realized:
\begin{equation}\label{eq25}
  3) \quad R/R_s \ll 1, \quad l/R_s \simeq 1, \quad R/l \ll 1.
\end{equation}
In this case, pore is situated close to granule boundary.
Let us note, that the case of small pores is distinguished by one more simplifying circumstance. It can be seen easily that healing of small pores
 $\frac{R(0)}{R_s(0)} \ll 1$ cannot be accompanied by a significant change of granule dimensions. Indeed, using the relation (\ref{eq19})(19), one can estimate an order of granule size change during the evolution. According to  (\ref{eq19}) this change can be written down in the form:
\[\frac{R_s(t)}{R_s(0)}=\sqrt[3]{1-\frac{R(0)^3}{R_s(0)^3}+\frac{R(t)^3}{R_s(0)^3}} \simeq 1-\frac{1}{3}\frac{R(0)^3}{R_s(0)^3}  \]
Hence, within small pore approximation, granule size does not change  $R_s(t) \approx R_s(0)=R_{0s}$ up to cubic order of smallness  $\frac{R(0)^3}{R_s(0)^3} $. Then, neglecting granule radius change, the equation system (\ref{eq21}) takes on a simpler form:
\begin{equation}\label{eq26}
\begin{cases}
  \frac{d L}{d \tau}=\frac{3\exp\left(\frac{A}{r}-\frac{B}{r^3}\right)}{r} \cdot \left[\frac{\alpha^2}{r^2} \cdot (\widetilde \Phi _1  + \widetilde \Phi _2 ) - \frac{\alpha}{r} \cdot \sqrt {1 + \frac{\alpha^2}{r^2}}  \cdot (\Phi _1  + \Phi _2 )\right]-\\
-\frac{6\exp\left(-\frac{A}{r_{s0}}\right)}{r}\cdot \left[\frac{\alpha^2}{r^2} \cdot \widetilde \Phi _2  - \frac{\alpha}{r} \cdot \sqrt {1 + \frac{\alpha^2}{r^2}}  \cdot \Phi _2 \right],\\
\frac{d r}{d \tau}=-\frac{\exp\left(\frac{A}{r}-\frac{B}{r^3}\right)}{r} \cdot \left[\frac{1}{2} + \frac{\alpha}{r} \cdot (\Phi _1  + \Phi _2 )\right]+\frac{2\exp\left(-\frac{A}{r_{s0}}\right)}{r}\cdot \frac{\alpha}{r} \cdot \Phi _2,\\
  r|_{\tau =0}=1,\\
	r_s|_{\tau =0}=r_{s0}, \\
  L|_{\tau =0}=\frac{l(0)}{R(0)}.
\end{cases}
\end{equation}
Let us now consider asymptotic case  (\ref{eq23}). By virtue of $L \gg r$, in Eqs. (\ref{eq26}), the expression for the parameter $\alpha$ is simplified
\begin{equation}\label{eq27}
\alpha \approx \frac{r_{s0}^2}{2L}\sqrt{\left(1-\frac{L^2}{r_{s0}^2}\right)^2}= \frac{r_{s0}^2}{{2L}}\left(1-\frac{L^2}{r_{s0}^2}\right),\end{equation}
and bispherical coordinates  $\eta_{1,2}$, that are defined according to(\ref{eq5}), are, correspondingly, equal to:
\begin{equation}\label{eq28}
 \eta _1  = \textrm{arsinh} \left( {\frac{r_s^2}{{2rL}}}\left(1-\frac{L^2}{r_s^2}\right) \right),\;\eta _2  = \textrm{arsinh} \left( \frac{r_s}{2L} \left(1-\frac{L^2}{r_s^2}\right) \right).\end{equation}
Since $\frac{\sinh\eta_1}{\sinh\eta_2}=\frac{r_s}{r} \gg 1$, then $\eta_1  \gg \eta_2$. In this case, series sums  can be estimated via following expressions:
\begin{equation}\label{eq29}
 \Phi_1 \approx \frac{1}{2\sinh2\eta_1},\; \Phi_2 \approx \frac{1}{\sinh2\eta_1},\;  \widetilde{\Phi}_1 \approx \frac{1+2\sinh^2\eta_1}{8\sinh^2\eta_1\cosh^2\eta_1},\; \widetilde{\Phi}_2 \approx \frac{\cosh\eta_1}{2\sinh^2\eta_1},$$
 $$ \sinh\eta_1=\frac{\alpha}{r},\quad \cosh\eta_1=\sqrt{1+\frac{\alpha^2}{r^2}}. \end{equation}
By substituting expressions (\ref{eq27})-(\ref{eq29}) into equation set  (\ref{eq26}), we find simplified equation set:

\begin{equation}\label{eq30}
\frac{d L}{d \tau}=-\frac{3}{2}\,\exp\left(\frac{A}{r}-\frac{B}{r^3}\right)\cdot\frac{r \left(\frac{L^2}{r_{s0}^2}\right)}{r_{s0}^2} ,\end{equation}
\begin{equation}\label{eq31}	
\frac{d r}{d \tau}=-\frac{\exp\left(\frac{A}{r}-\frac{B}{r^3}\right)}{r} \cdot \left[1 + \frac{r}{2L}\cdot\left(\frac{L^2}{r_{s0}^2}\right) \right]+\frac{\exp\left(-\frac{A}{r_{s0}}\right)}{r}
\end{equation}
Equations (\ref{eq30})-(\ref{eq31}) are written down up to  $L^2/r_{s0}^2$ terms. This nonlinear set signifies that, at high gas concentration, pore size increases monotonously while moving towards the granule center. Besides,  smallness of the right part of (\ref{eq30}) means that pore displacement towards granule center during  characteristic time of the establishment of stationary pore radius is small.

It is interesting to compare the behavior of the pore in this asymptotic mode with the solutions of complete equation set (\ref{eq26}). In Fig. \ref{fg3} the numerical solutions of exact (\ref{eq26}) and approximate  (\ref{eq30})-(\ref{eq31}) equation sets are shown with the same initial conditions  $r|_{\tau =0}=1$, $r_s|_{\tau =0}=100$, $L|_{\tau =0}=10$ , $A=10^{-1}$ and different values of parameter $B$, connected to the value of gas concentration $N_g$: $B=0.25$  $(N_{g}=1.05\cdot 10^{5})$,  $B=0.10101$ $(N_{g}=3.17 \cdot 10^{4})$, $B=0.025$ $(N_{g}=1.05 \cdot 10^{4})$. Fig. \ref{fg3} 2 demonstrates good agreement of the approximate solution with the solution of the complete equation set for pore radius time change for various gas concentrations.  In Fig. \ref{fg3}  the plots are also shown for the time change of center-to-center distance between the pore and the granule. These plots demonstrate, that at "high" gas concentrations, in the case of approximate solution, pore displacement speed is most underestimated   as compared to other modes.

Such good agreement allows us to consider pore radius change at zeroth-order  at  $L^2/r_{s0}^2 \ll 1$. In this case we obtain a simple equation for the radius of an immobile pore:
\begin{equation}\label{eq32}
\frac{d r}{d \tau}=-\frac{\exp\left(\frac{A}{r}-\frac{B}{r^3}\right)}{r}+\frac{\exp\left(-\frac{A}{r_{s0}}\right)}{r}\end{equation}
It can be seen from here that the sign of right part of Eq. (\ref{eq32}) depends on the gas parameter $B$. One can easily obtain general solution of Eq.  (\ref{eq32}) in integral form
\begin{equation}\label{eq33} \tau+\textrm{const}= e^{A/r_{s0}}\int\frac{rdr}{1-e^{A/r-B/r^3+A/r_{s0}}} \end{equation}
\begin{figure}
  \centering
  \includegraphics[ height=5.3 cm]{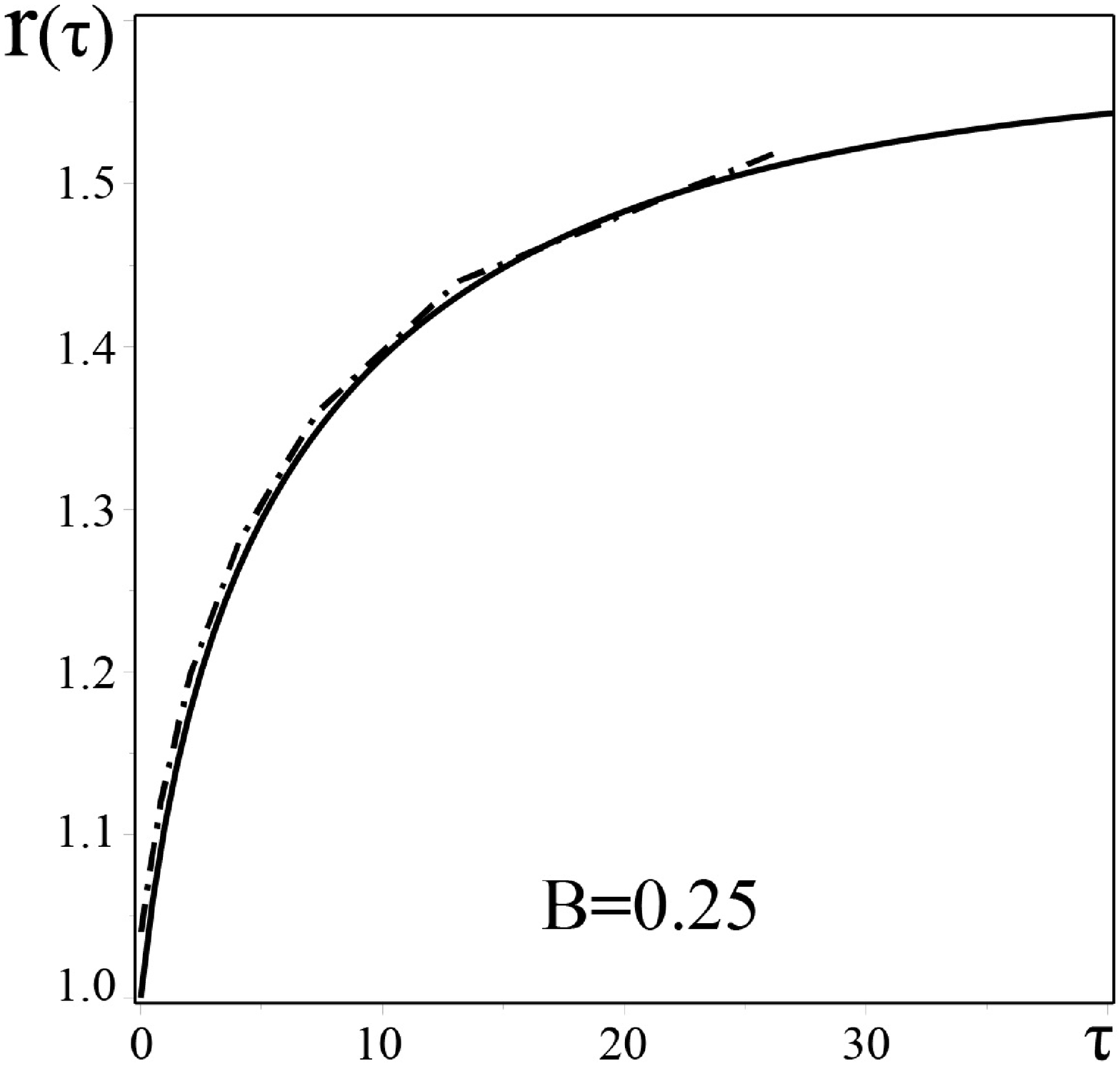}
	 \includegraphics[ height=5.3 cm]{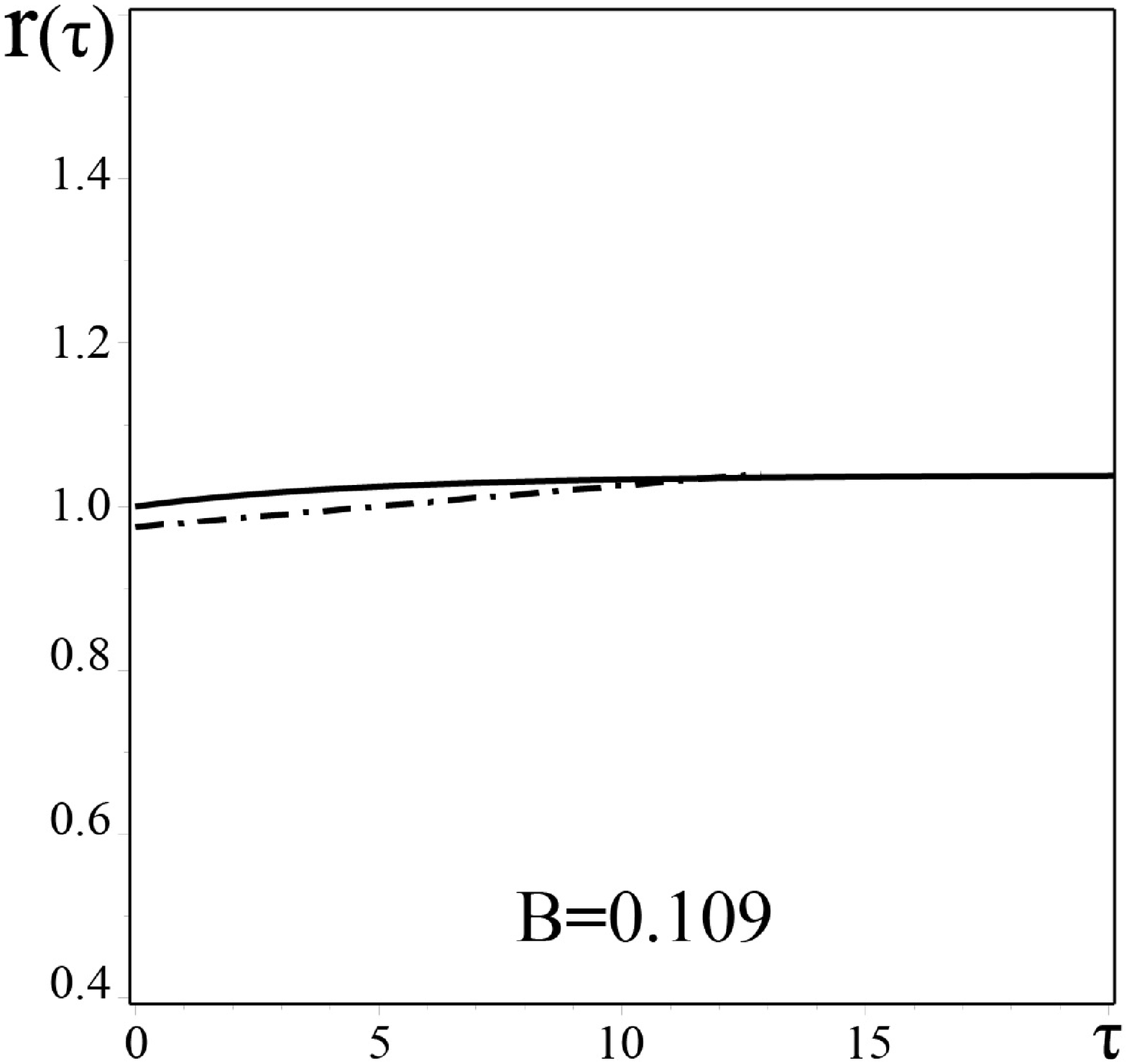}
	     \includegraphics[ height=5.3 cm]{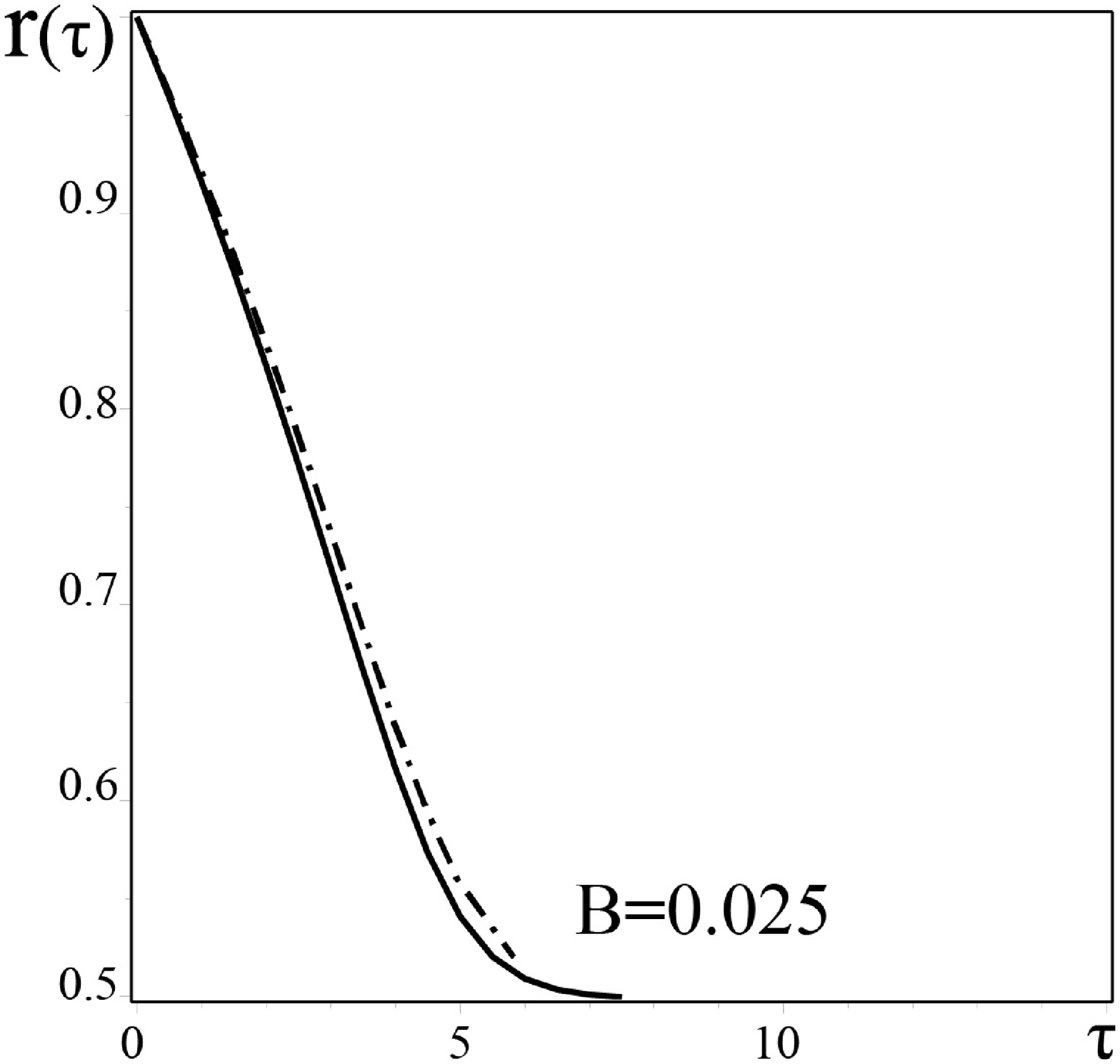}\\
  \caption{In Fig. solid line  numerical  solution equation set (\ref{eq26}),  dash-and-dot -- analytical solution of Eq. (\ref{eq32})  for initial conditions $r|_{\tau =0}=1$, $r_s|_{\tau =0}=100$, $L|_{\tau =0}=10$, $A=10^{-1}$ and different values of gas parameter $B$.}
    \label{fg4}
\end{figure}
Here $\textrm{const}$ is determined by initial conditions.

Let us consider the following evolution stage of "small" ($r \ll r_{s0}$)gas-filled pores where $A \ll r$  and $B \ll r^3$: $$ \tau \approx -\frac{\exp\left(\frac{A}{r_{s0}}\right)}{A}\left(\frac{r(\tau)^3}{3}-\frac{r(0)^3}{3}+\left(\frac{B}{A}\right)\cdot \left(r(\tau)-r(0)\right)- \right.$$
\begin{equation}\label{eq34}   \end{equation}
$$\left.-\frac{1}{2}\left(\frac{B}{A}\right)^{3/2}\cdot \ln \frac{\left(\sqrt{B/A}+r(\tau)\right)\left(\sqrt{B/A}-r(0)\right)}{\left(\sqrt{B/A}-r(\tau)\right)\left(\sqrt{B/A}+r(0)\right)} \right) $$
In the absence of gas inside the pore $B=0$, this dependence coincides with that obtained in \cite{21s}, where it was shown that vacancy pore healing time is proportional to third power if initial pore radius  $r(0)$ and to material temperature $T$, since $A \sim 1/T$ (of course, without taking into account temperature dependence of diffusion coefficient). In the presence of the gas $B\neq 0$ one can consider the following modes:
\begin{enumerate}
  \item  gas density is "large" $B \gg A$;
  \item  "equilibrium" gas concentration  $B \cong A$, at which pore radius practically does not change;
  \item "low" gas density $B\ll A$.
\end{enumerate}
In Fig. \ref{fg4} dash line indicates plots of analytical solutions (\ref{eq34}) at different values of parameter $B$ that correspond to the described above modes. Solid line in Fig. \ref{fg4} indicates the numerical solution of equation set (\ref{eq26}). Here we observe a good agreement between analytical and numerical solutions.

Let us now turn to the case  (\ref{eq24}) of a small pore situated close to the granule center:
\begin{equation}\label{eq35}
 R/R_s \ll 1,\quad l/R_s \ll 1,\quad  R \gg l. \end{equation}
Such inequalities comply with the geometrical condition   $R/R_s+l/R_s\leq 1$. Taking into account  (\ref{eq35}), it is easy to find expressions for parameter $\alpha$ and bispherical coordinates $\eta_{1,2}$:
\begin{equation}\label{eq36} \alpha \approx \frac{r_s^2}{2L}\left(1- \frac{r^2}{r_{s0}^2}\right),\; \eta _1  \approx \textrm{arsinh} \left( {\frac{r_{s0}^2}{{2rL}}}\left(1-\frac{r^2}{r_{s0}^2}\right) \right),\;\eta _2  \approx \textrm{arsinh} \left( \frac{r_s}{2L} \left(1-\frac{r^2}{r_{s0}^2}\right) \right)
\end{equation}
 It can be seen from here, that, for small pores, the relation  $\eta_1 \gg \eta_2 $ is valid. Using the estimation of series sums by formulas (\ref{eq29}), we approximate pore evolution equations for such case.

\begin{figure}
  \centering
  \includegraphics[ height=7 cm]{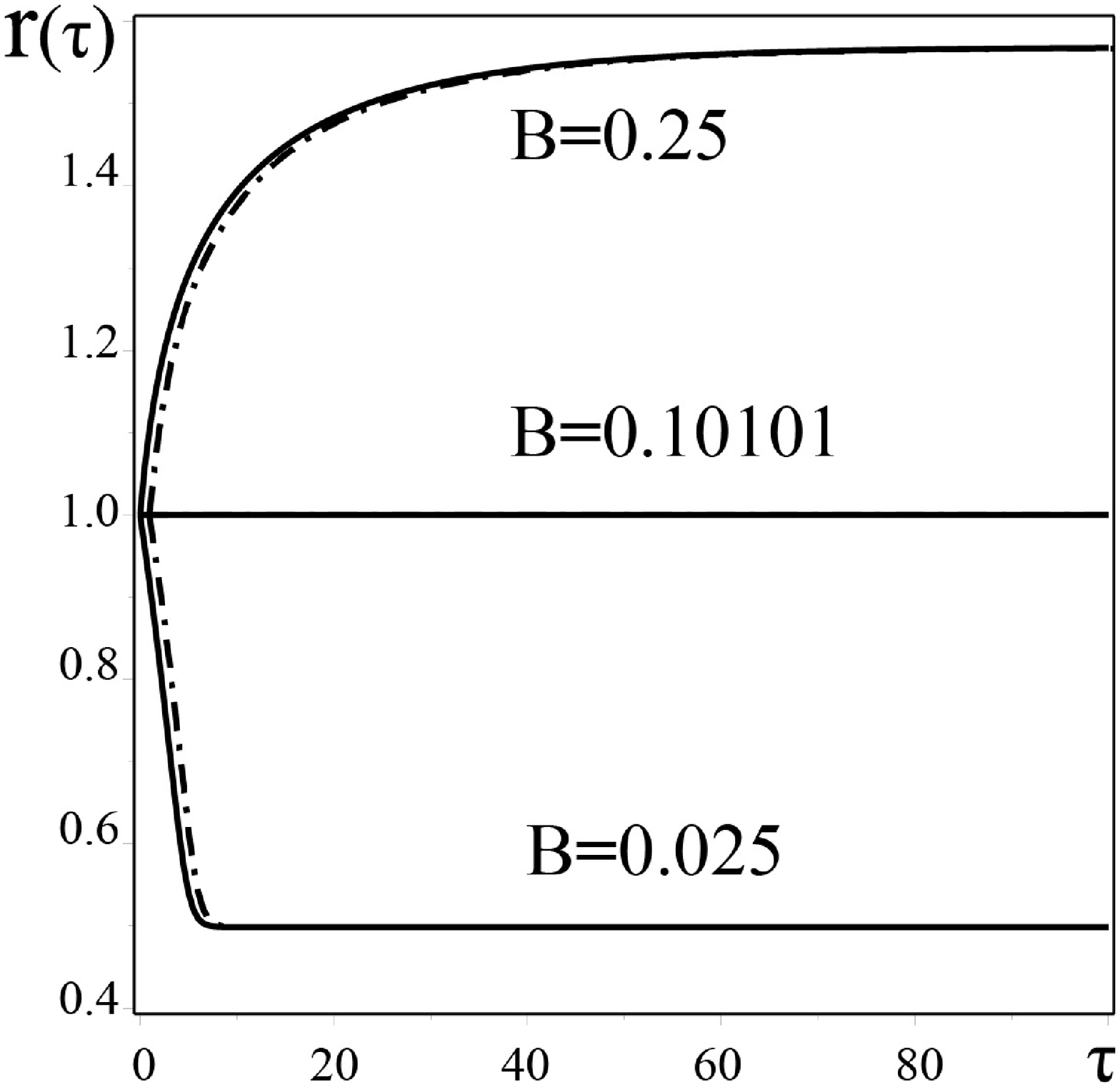}
	\includegraphics[ height=7 cm]{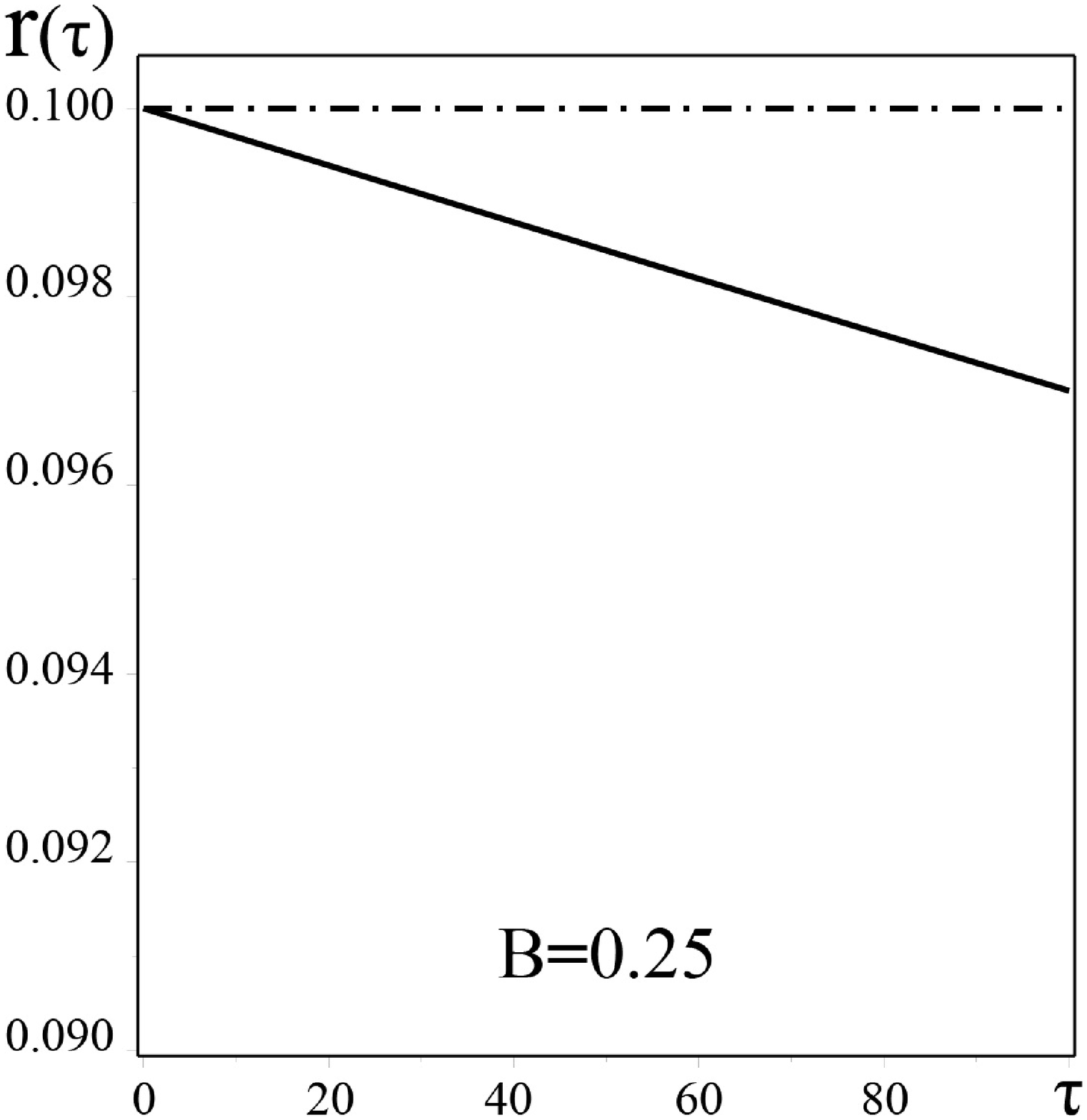}\\
	\includegraphics[ height=7 cm]{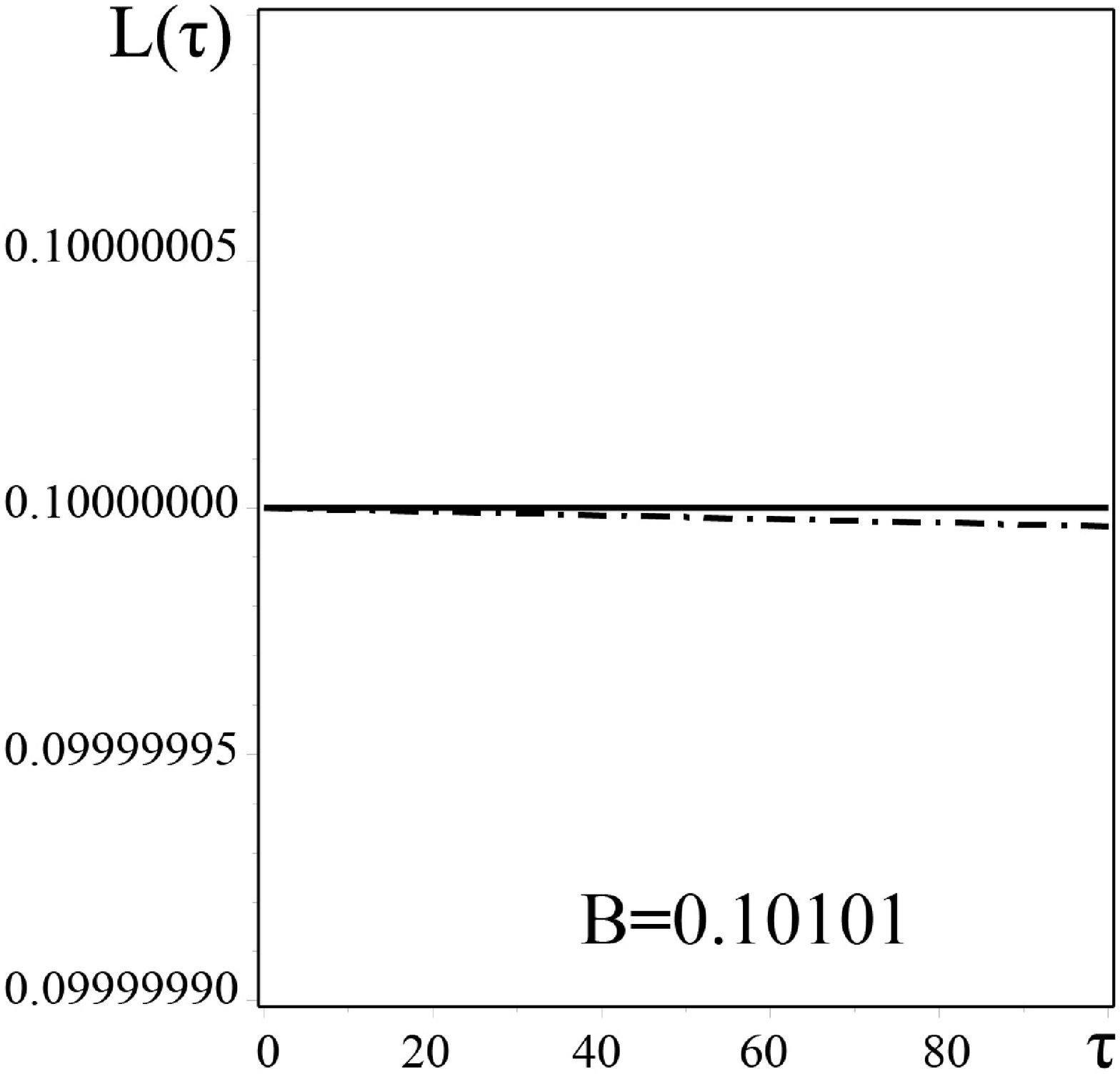}
	\includegraphics[ height=7 cm]{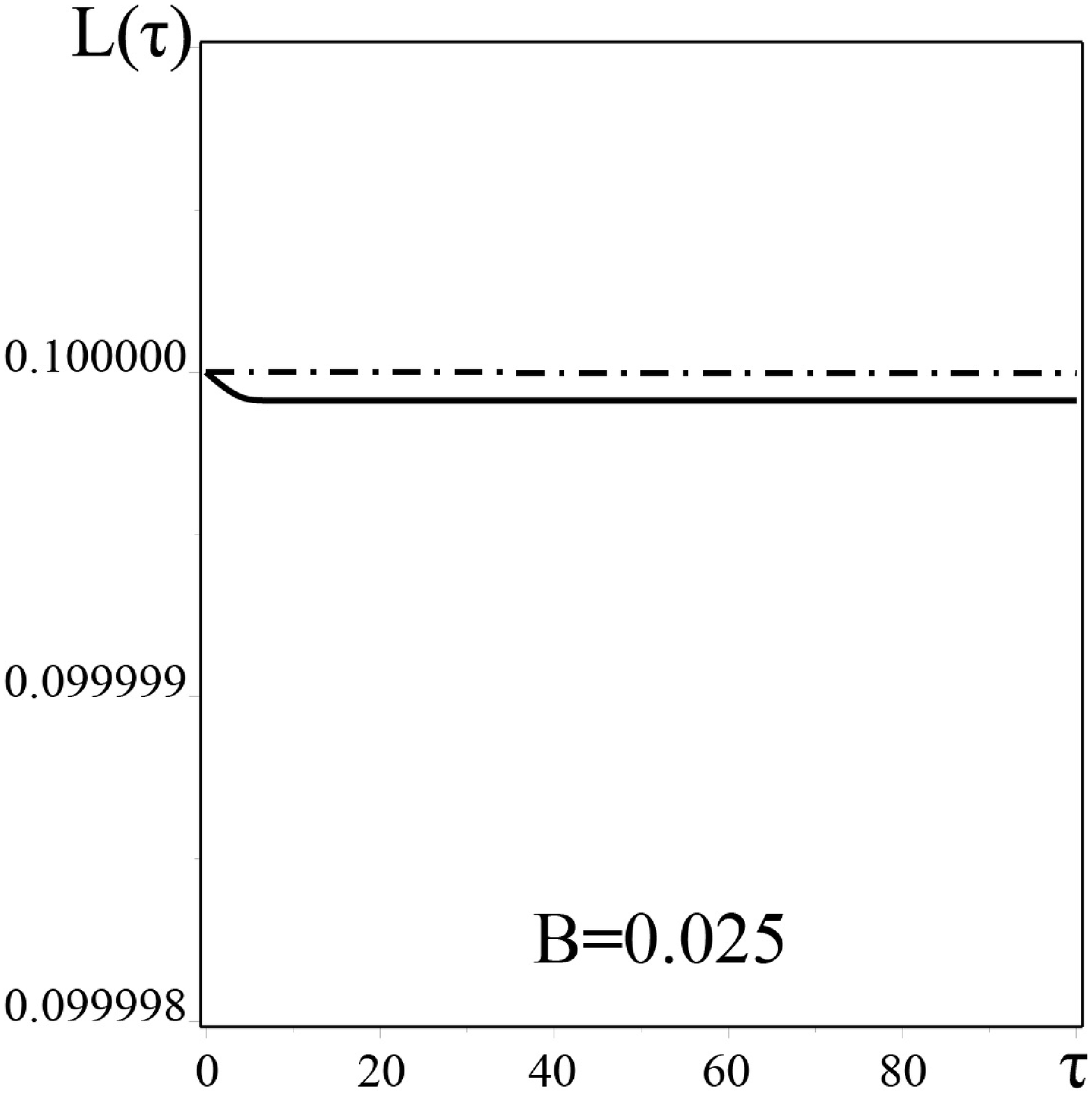}\\
  \caption{{ On  the upper left, the dependence is shown of  pore radius $r$ on time $\tau$ for different values of gas parameter $B$: $B=0.25$, $B=0.10101$  and $B=0.025$. Solid line corresponds to the numerical solution of complete equation set (\ref{eq26}), dash-and-dot line corresponds to the numerical solution of approximate equations (\ref{eq37})-(\ref{eq38});  on the upper right the dependence is shown of distance $L$ on time $\tau$ for parameter $B=0.25$; below on the left, the dependence is shown of distance time-change $L$ on time $\tau$ for parameter for parameter $B=0.10101$; below on the left, the dependence is shown of distance time-change $L$ on time $\tau$ for parameter for parameter $B=0.025$. All are solutions obtained at initial conditions $r|_{t=0}=1$, $r_s|_{t=0}=100$, $L|_{t=0}=0.1$ and $A=10^{-1}$.}}
\label{fg5}
\end{figure}

 \begin{equation}\label{eq37} \frac{d L}{d \tau}=-\frac{3}{2}\cdot \exp\left(\frac{A}{r}-\frac{B}{r^3}\right)\cdot \frac{r\left(\frac{L}{r_{s0}}\right)^2}{r_{s0}^2 \left(1-\frac{r^2}{r_{s0}^2}\right)^2} \end{equation}
\begin{equation}\label{eq38} \frac{d r}{d \tau}=-\frac{\exp\left(\frac{A}{r}-\frac{B}{r^3}\right)}{r}\cdot \left[1+\frac{1}{2}\cdot \frac{rL}{r_{s0}^2\left(1-\frac{r^2}{r_{s0}^2}\right)}\right]+\frac{\exp\left(-\frac{A}{r_{s0}}\right)}{r} \end{equation}

\begin{figure}
  \centering
  \includegraphics[ height=5.3 cm]{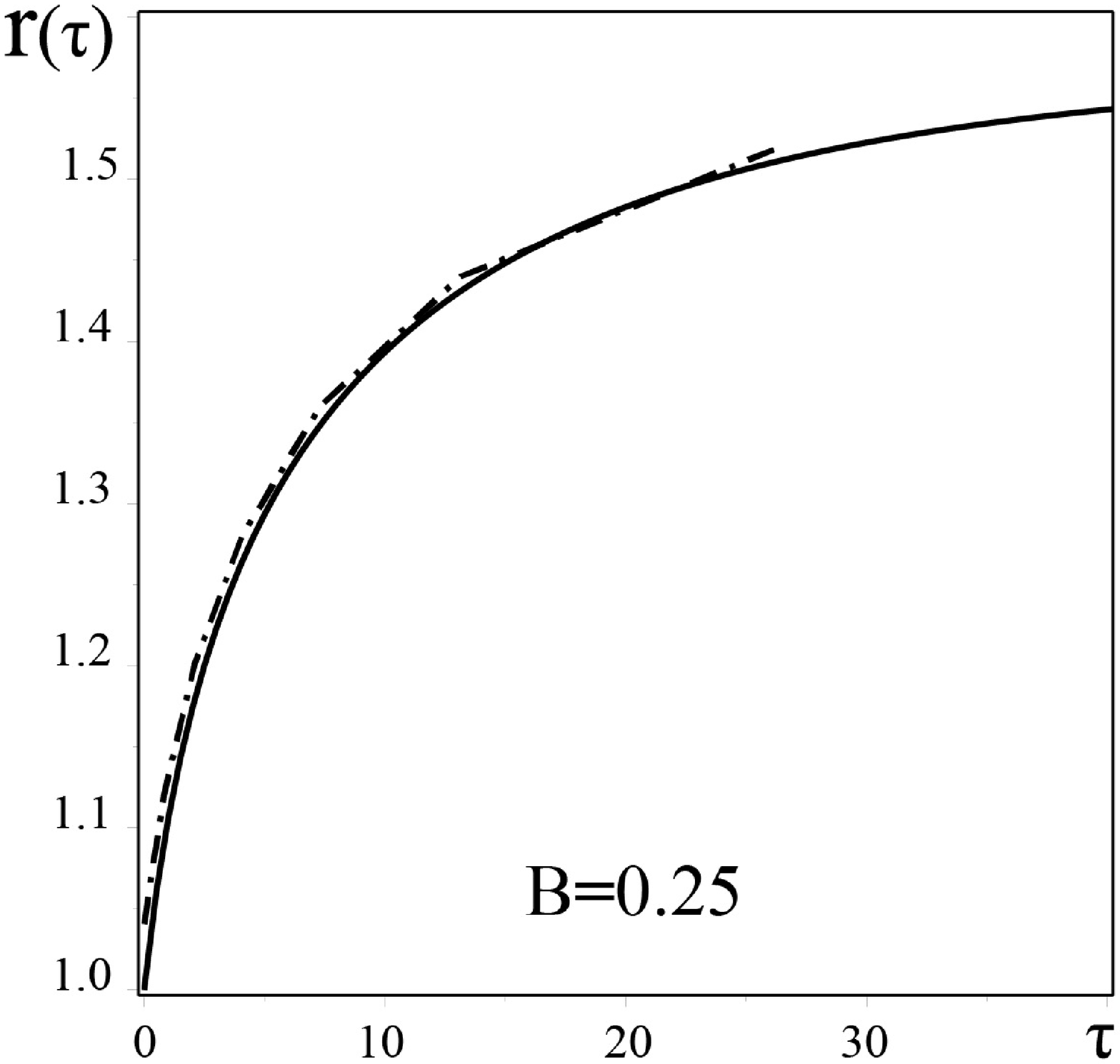}
	 \includegraphics[ height=5.3 cm]{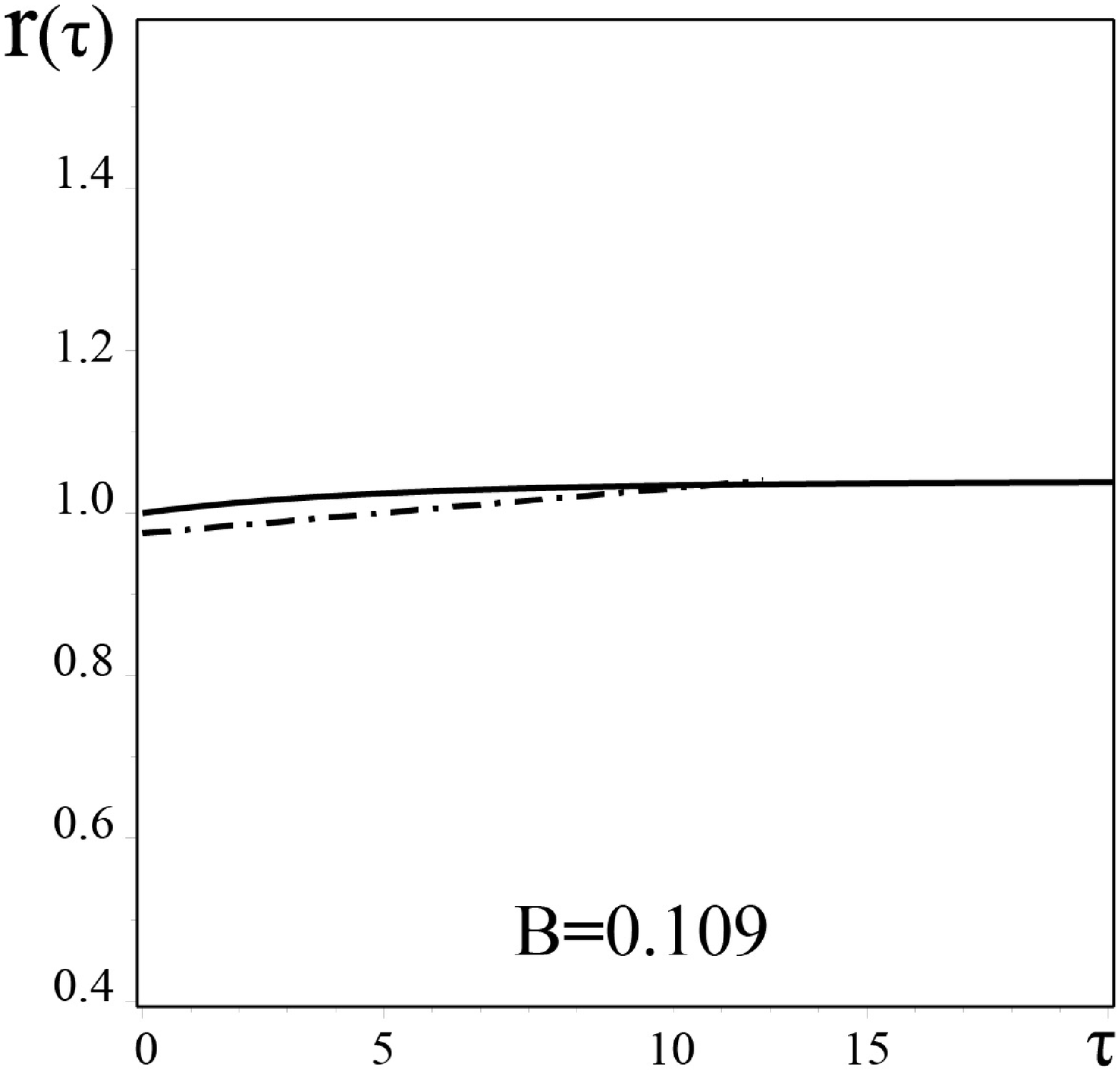}
	     \includegraphics[ height=5.3 cm]{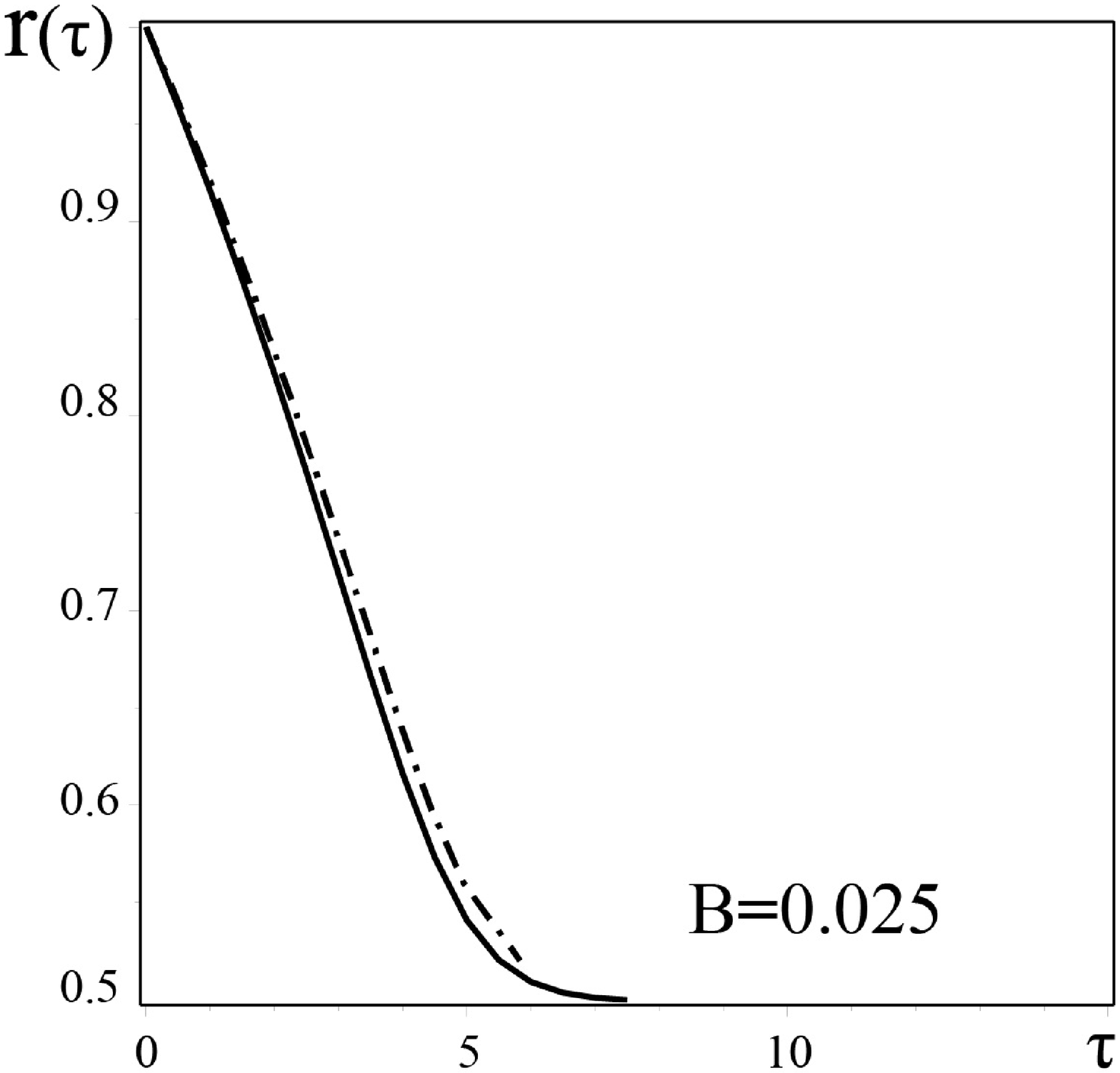}\\
  \caption{In Fig. solid line  numerical  solution  equation set (\ref{eq26}),  dash-and-dot   -- analytical solution of Eq. (\ref{eq38})  for initial conditions $r|_{\tau =0}=1$, $r_s|_{\tau =0}=100$, $L|_{\tau =0}=0.1$, $A=10^{-1}$ and different values of gas parameter $B$.}
    \label{fg6}
\end{figure}

In the Fig. \ref{fg5} numerical solutions of Eqs. (\ref{eq26}) (solid line) and Eqs.  (\ref{eq37})-(\ref{eq38}) (dashed line) are shown for initial conditions, satisfying inequalities (\ref{eq35}): $r|_{\tau=0}=1$, $r_s|_{\tau=0}=100$, $L|_{\tau=0}=0.1$ and $A=10^{-1}$ for different variants of gas concentration: 1) $B=0.25$  $(N_{g}=1.05\cdot 10^{5})$, 2)  $B=0.10101$ $(N_{g}=3.17 \cdot 10^{4})$ and 3) $B=0.025$ $(N_{g}=1.05 \cdot 10^{4})$. The upper left part of Fig. \ref{fg5} demonstrates a good agreement of numerical solutions of Eqs. (\ref{eq25}) and (\ref{eq37})-(\ref{eq38}) for pore radius change. In the right upper part of Fig. \ref{fg5} time change of center-to-center distance between the pore and the granule is demonstrated, correspondingly, for equation set (\ref{eq26}) and Eqs. (\ref{eq37})-(\ref{eq38}) at $B=0.25$. The lower left part of Fig. \ref{fg5} shows time change of center-to-center distance between the pore and the granule correspondingly for equation set (\ref{eq26}) and Eqs. (\ref{eq37})-(\ref{eq38}) at $B=0.10101$. Lower right part of Fig. \ref{fg5} shows time change of center-to-center distance between the pore and the granule correspondingly for equation set (\ref{eq26}) and  Eqs. (\ref{eq37})-(\ref{eq38}) at $B=0.025$. Similarly to the previous case, the pore is almost immobile: $L(t) \approx L(0)$. Therefore, we can confine ourselves to zeroth approximation for the pore evolution analysis. In this case,  (\ref{eq31}) is obtained. The analytical solution of this equation well agrees with the numerical solution of equation set  (\ref{eq26}) for initial conditions  $r|_{\tau=0}=1$, $r_s|_{\tau=0}=100$, $L|_{\tau=0}=0.1$, $A=10^{-1}$ at various parameters $B$:  $B=0.25$,  $B=0.10101$ and $B=0.025$.  These solutions are shown in Figs. \ref{fg6}.

Let us, finally, turn to the discussion of the mode (\ref{eq25}), when the pore is situated close to the granule boundary. In this case, the relation $l/R_s$ is close to unity:
\[\frac{l}{R_s}=1-\varepsilon,\]
Here $\varepsilon$ is small parameter, on which the asymptotic expansion is conducted.

\begin{figure}
  \centering
  \includegraphics[ height=7 cm]{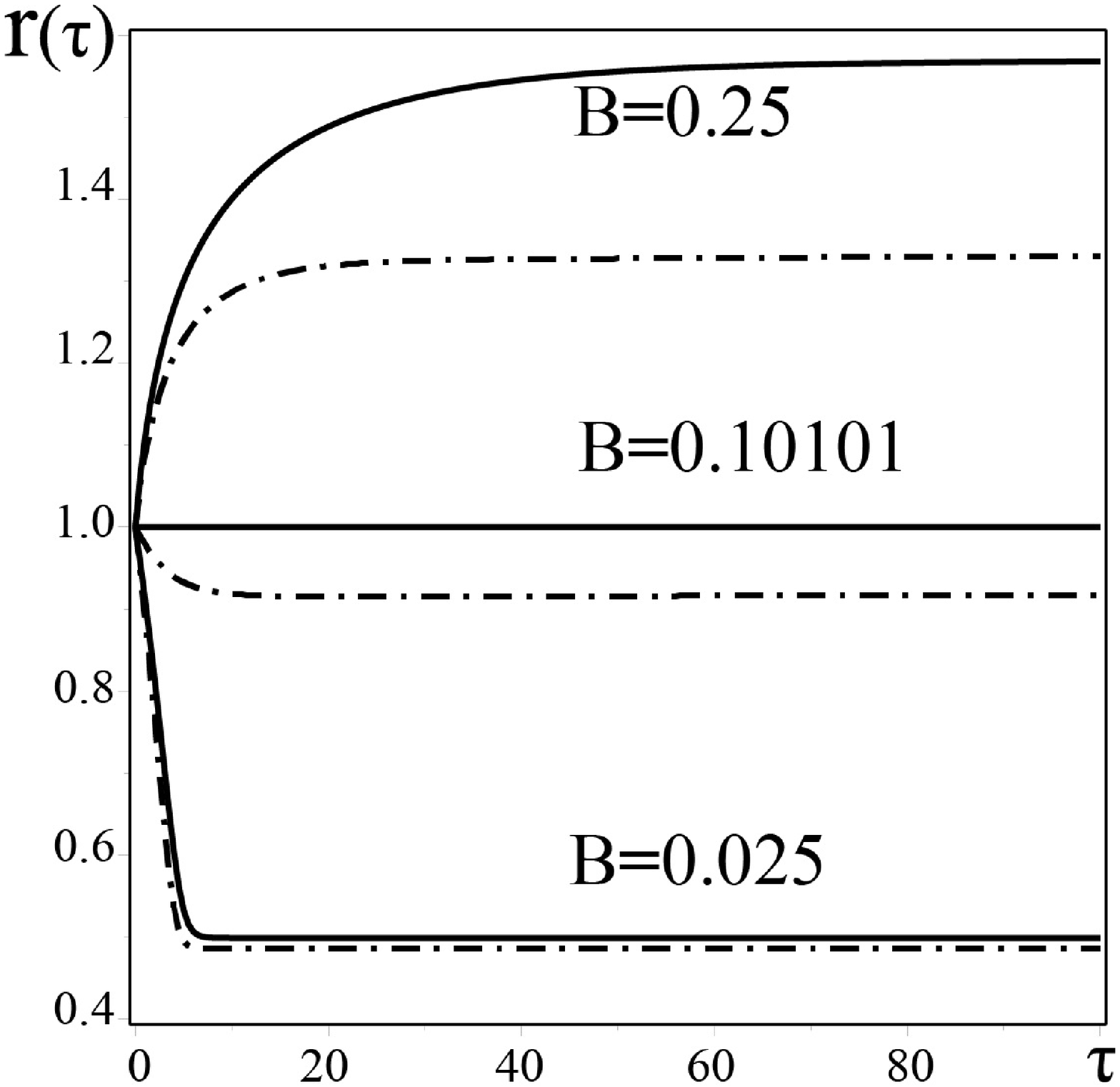}
	\includegraphics[ height=7 cm]{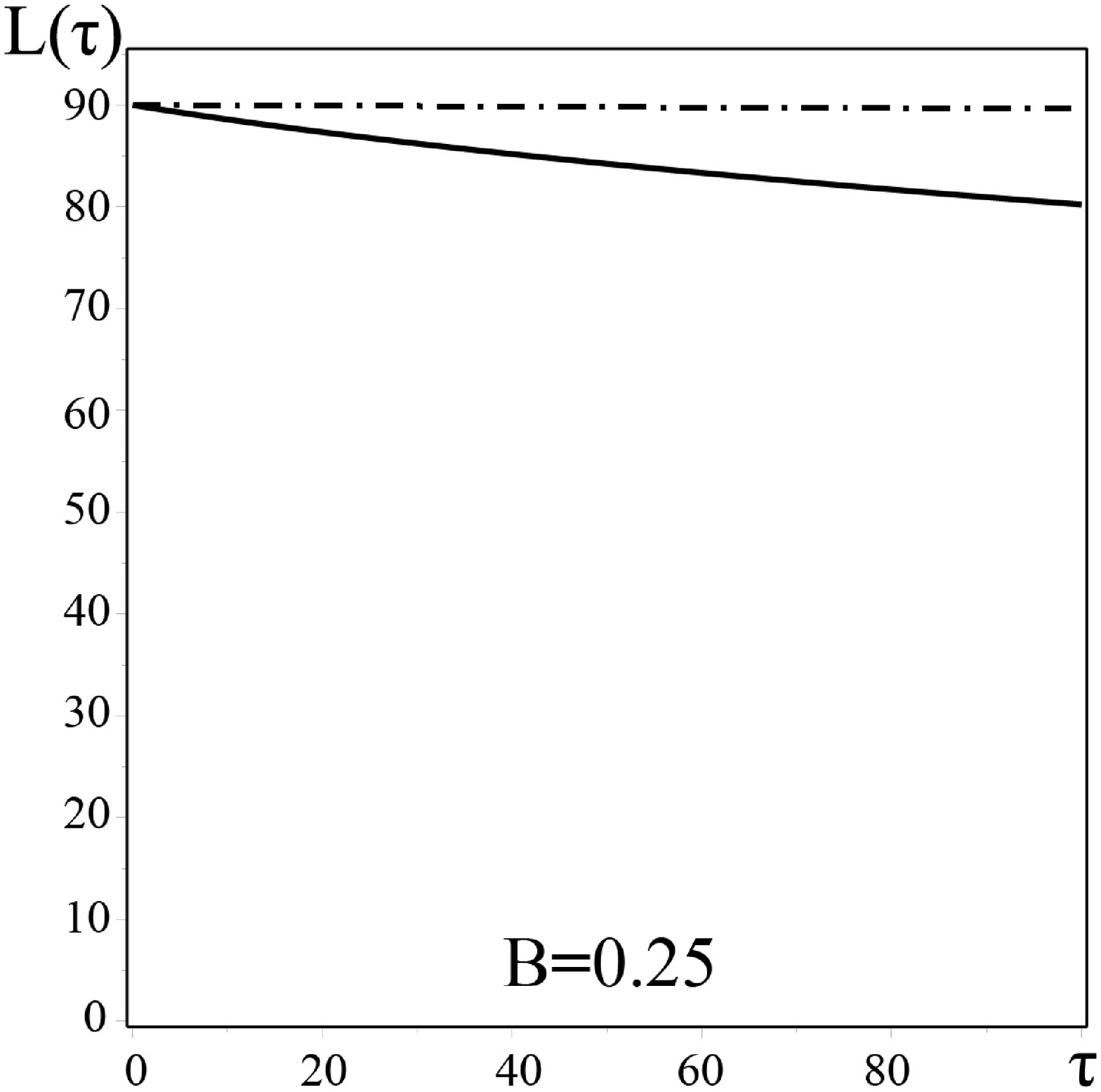}\\
	\includegraphics[ height=7 cm]{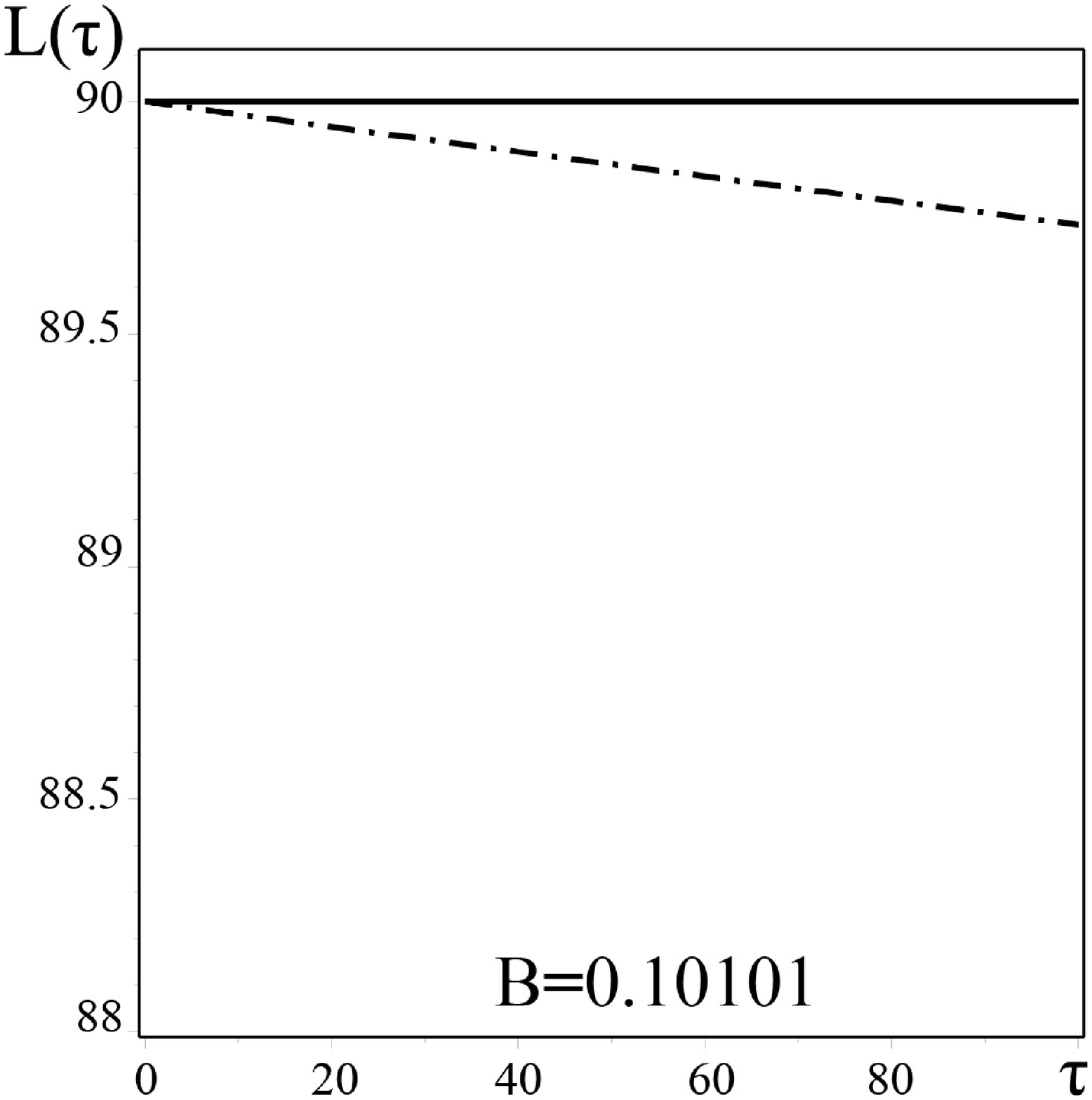}
	\includegraphics[ height=7 cm]{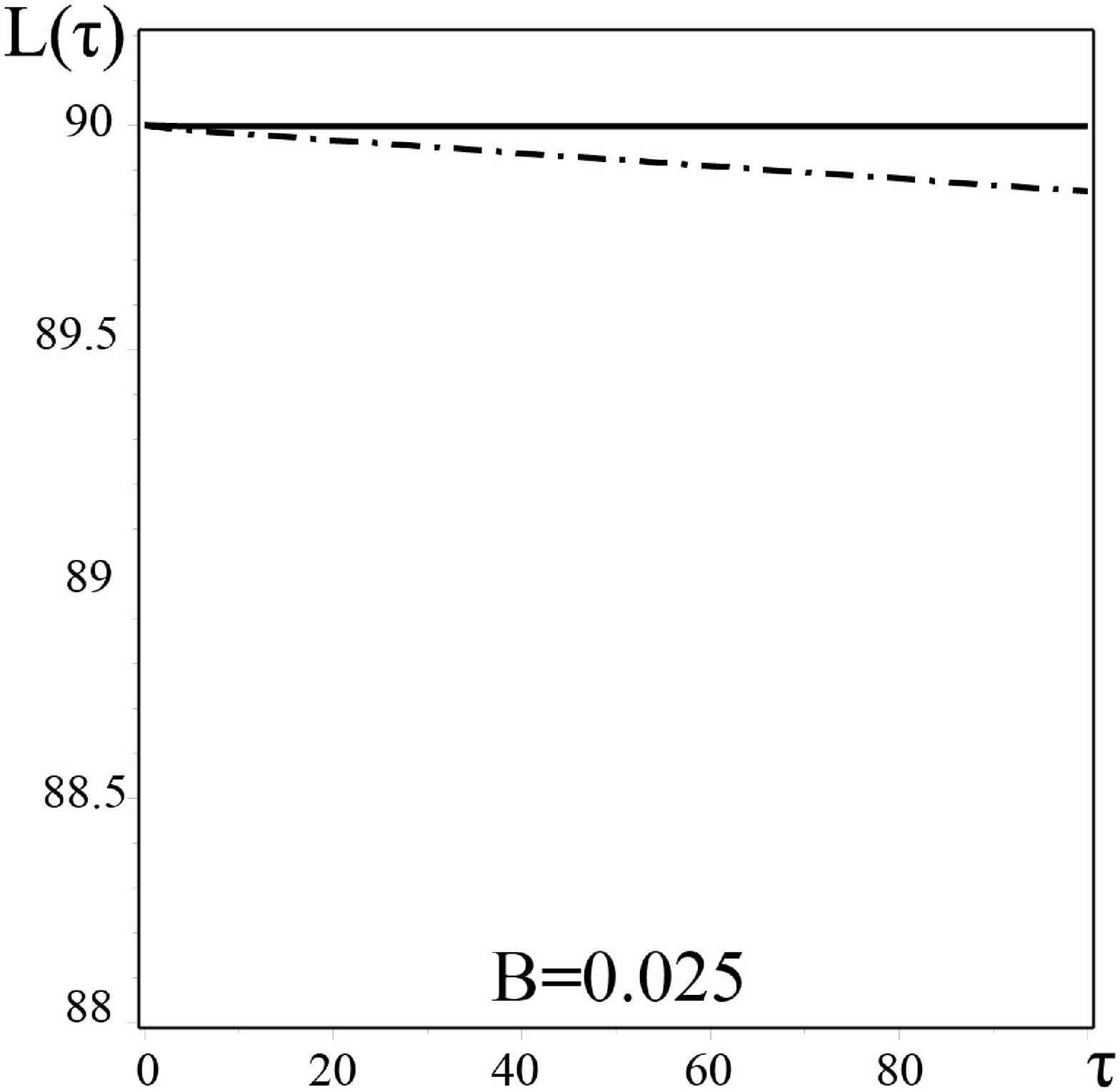}\\
  \caption{On the  upper left, dependencies of pore radius  $r$ on time $\tau$ are given for different values of gas parameter $B$: $B=0.25$, $B=0.10101$  and $B=0.025$. Solid line corresponds to the numerical solution of complete equation set (\ref{eq26}), dash-and-dot line corresponds to the numerical solution of approximate equations (\ref{eq37})-(\ref{eq38});  on the upper right, dependence is shown of distance $L$ on time $\tau$ for parameter $B=0.25$; below on the left, dependence is shown of   distance time-change $L$ on time $\tau$ for parameter $B=0.10101$; below on the right, dependence is shown of   distance time-change $L$ on time $\tau$ for parameter $B=0.025$. All solutions have been obtained at initial conditions $r|_{t=0}=1$, $r_s|_{t=0}=100$, $L|_{t=0}=90$ and $A=10^{-1}$.}
\label{fg7}
\end{figure}

Parameter  $\varepsilon$ value is restricted by the geometrical inequality (the pore inside the granule)
\[\frac{R}{R_s} \leq \varepsilon .\]
In asymptotic expansion we will take into account the terms of the order of  $\varepsilon^2$. With account of this remark, parameter $\alpha$ and, correspondingly, bispherical coordinates  $\eta_{1,2}$ obtained within the small pore approximation $R\ll R_s$ and $R \ll l$, take on the form:
\begin{equation}\label{eq39}
 \alpha \approx \frac{r_s^2}{2L}\varepsilon(2-\varepsilon), \; \eta_1 \approx \ln \left(\frac{r_s}{r}(2\varepsilon+\varepsilon^2)\right),\; \eta_2 \approx \varepsilon + \frac{\varepsilon^2}{2}. \end{equation}
It can be seen from here, that $\eta_1 \gg \eta_2$ , therefore we can use previous estimates for the sums of series given by formulas (\ref{eq29}). Substituting  (\ref{eq29}) and (\ref{eq39}) into the right part of Eq. (\ref{eq26}), we obtain pore evolution equations within approximation  (\ref{eq25}):
 \begin{equation}\label{eq37a} \frac{d L}{d \tau}=-\frac{3}{8}\cdot \exp\left(\frac{A}{r}-\frac{B}{r^3}\right)\cdot \frac{r\left(\frac{L}{r_{s0}}\right)^2}{r_{s0}^2 \left(1-\frac{L}{r_{s0}}\right)^2} \end{equation}
\begin{equation}\label{eq38a} \frac{d r}{d \tau}=-\frac{\exp\left(\frac{A}{r}-\frac{B}{r^3}\right)}{r}\cdot \left[1+\frac{1}{2}\cdot \frac{rL}{r_{s0}^2\left(1-\frac{L^2}{r_{s0}^2}\right)}\right]+\frac{\exp\left(-\frac{A}{r_{s0}}\right)}{r} \end{equation}
In Fig. \ref{fg7}, the numerical solutions are shown both of the exact equation set (\ref{eq26}) and of the approximate one (\ref{eq37a})-(\ref{eq38a}) with the same initial conditions  $r|_{\tau =0}=1$, $r_s|_{\tau =0}=100$, $L|_{\tau =0}=90$ and $A=10^{-1}$. The left upper part of Fig. \ref{fg7} demonstrates very good agreement of the time dependences of pore radius. In Fig. \ref{fg7} the plots are shown for time dependence of the center-to-center distance between the pore and the granule. It can be seen from the figure, that the displacement of the pore towards the granule center, obtained from exact equation set (\ref{eq26}) exceeds that observed in  approximate equation set (\ref{eq37a})-(\ref{eq38a}).

\subsection{Large pores}

Let us now proceed to discussing the evolution of large pores. Let us begin with the notion, that asymptotic mode
\begin{equation}\label{eq40}
 R/R_s \cong 1,\quad l/R_s \cong 1,\quad  R/l \cong 1. \end{equation}
is not, in fact, realized. Indeed, let us take into account the closeness of the two firs relations to the unity
\begin{equation}\label{eq41}
\frac{R}{R_s}=1-\varepsilon_1,\quad \frac{l}{R_s}=1-\varepsilon_2, \end{equation}
where $\varepsilon_1 \ll 1$ and $\varepsilon_2 \ll 1$ are small parameters. Substituting  (\ref{eq41}) into geometrical condition (\ref{eq22}), we find  $1 \leq \varepsilon_1+\varepsilon_2$. Since $\varepsilon_{1,2}$ are small parameters, this inequality does not hold. Thus, mode (\ref{eq40}) is not compatible with geometrical condition (\ref{eq22}).

Let us consider the valid regime of large pore evolution when relations between values  $R$, $R_s$, $l$ are the following:
\begin{equation}\label{eq42}
 4) \quad R/R_s \cong 1,\quad l/R_s \ll 1,\quad  R \gg l. \end{equation}
Let us write down the first relation as  $R/R_s=1-\epsilon$, where $\epsilon$ is a small parameter of asymptotic expansion. With an account of the validity of conservation low for the volume of granule material, we can find, from Eq. (\ref{eq19}) granule radius change
 \[R_s(t)=\left(R_s(0)^3-R(0)^3+R(t)^3 \right)^{1/3} \]
or, in dimensionless units,
\begin{equation}\label{eq43} r_s(t)=\left(r_s(0)^3-r(0)^3+r(t)^3 \right)^{1/3} \end{equation}
Using this relation, we can describe the large pore evolution by the following dimensionless equations (\ref{eq21}). The numerical solution of equation set
\begin{figure}
  \centering
  \includegraphics[ height=7 cm]{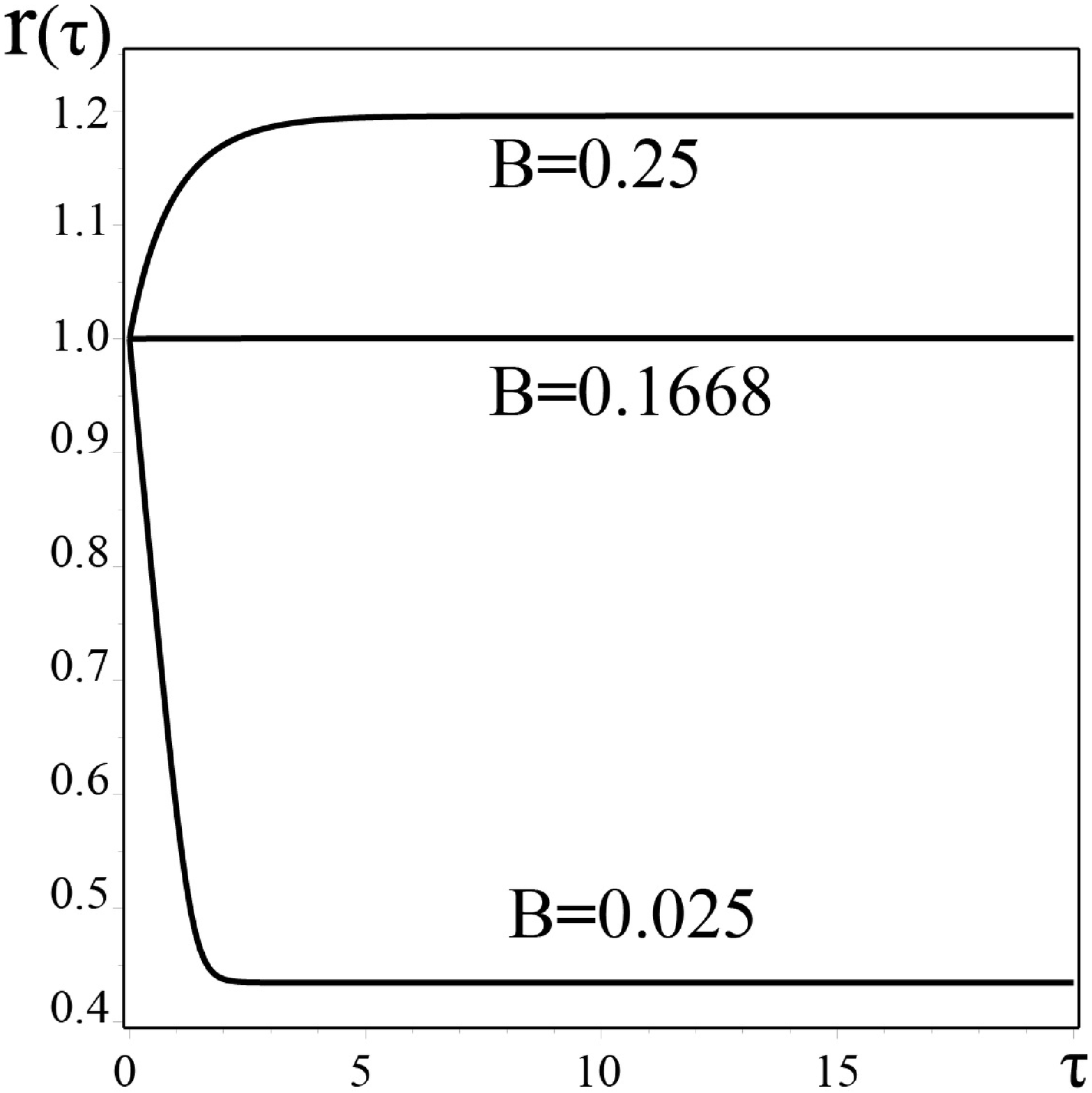}
	\includegraphics[ height=7 cm]{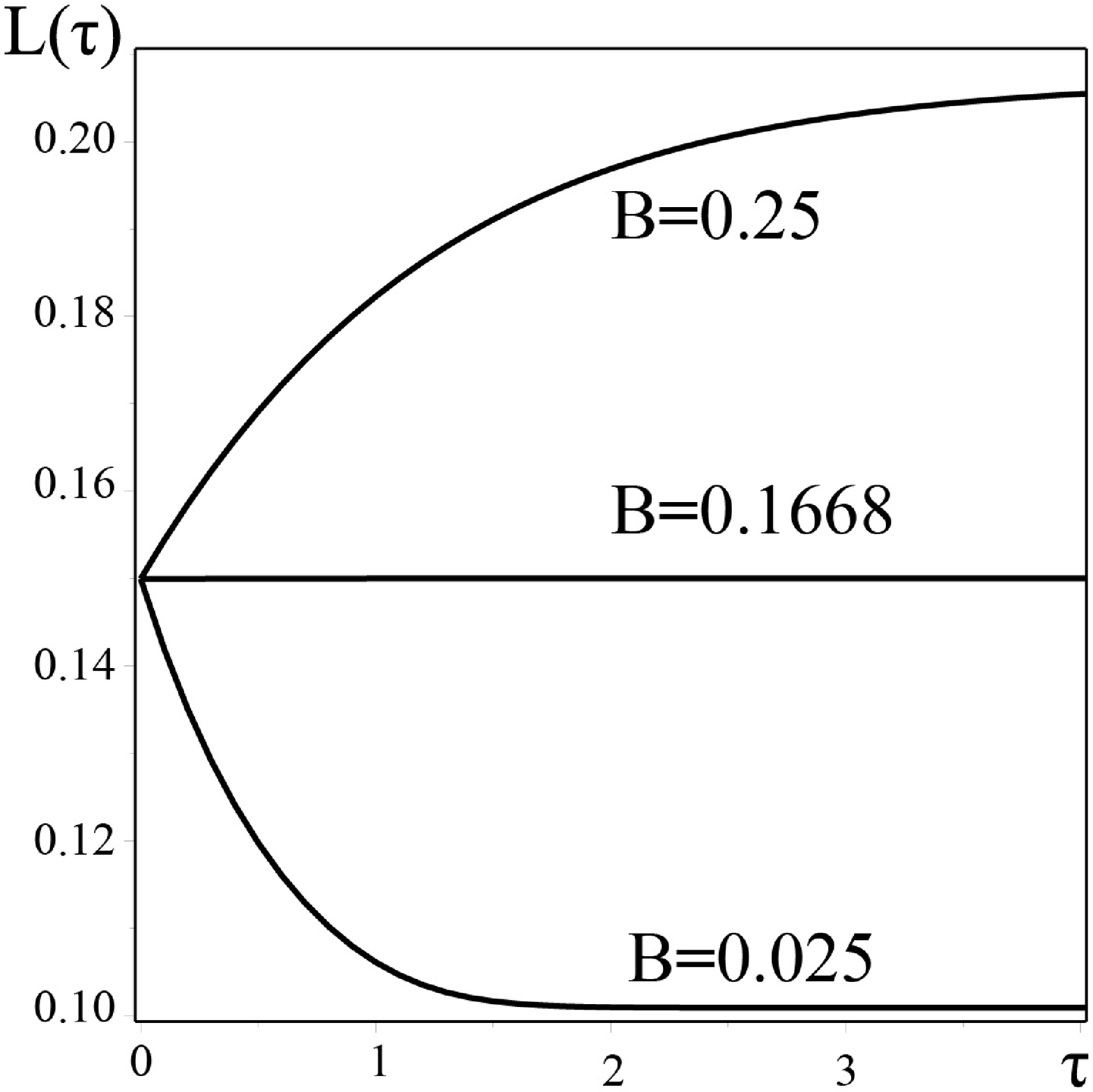}\\
\caption{Dependencies are shown for the case of large pore of radius  $r$ on time $\tau$ (on the left) and of distance $L$ on time $\tau$ (on the right)  for different values of gas parameter $B$: $B=0.25$, $B=0.1668$ and $B=0.025$. All plots correspond to numerical solutions of equation set (\ref{eq21}) for initial conditions $r|_{t=0}=1$, $r_s|_{t=0}=1.5$, $L|_{t=0}=0.15$ and $A=10^{-1}$.}
\label{fg8}
\end{figure}
is shown in Fig. \ref{fg8} for initial conditions $r|_{\tau =0}=1$, $r_s|_{\tau =0}=1.5$, $L|_{\tau=0}=0.15$, $A=10^{-1}$ at different values of gas parameter $B=0.25$, $B=0.1 668$ and $B=0.025$. Fixed values of gas parameter $B$ a chosen in such a way that  three characteristic evolution modes of large pore could be demonstrated. The first mode corresponds to the case of "high" gas density inside the pore ($B=0.25$ or $N_{g}=1.05\cdot 10^{5}$). It can be seen from the plot at $B=0.25$ , that large pore evolution is accompanied by an increase of pore radius up to some stationary value $r_{cr}^{h}$ when pore is shifted relative to granule center at some critical distance $L_{cr}^{h}$. Second evolution mode is the case of the pore "at rest", i.e. at some definite  value of gas parameter $B=0.1668$, radius and position of the pore do not change. Third case relates to small gas concentration. Finally, third evolution mode pore  corresponds to "small" gas pressure ($B=0.025$ or $N_{g}=1.05 \cdot 10^{4}$), at which  decrease of the pore radius down to some stationary value  $r_{cr}^{l}$ occurs, that is accompanied by pore shifting towards granule center at some critical distance $L_{cr}^{l}$ (see Fig. \ref{fg8}).

Let us consider asymptotic mode (\ref{eq42}) for a large pore, confining ourselves, in connection with (\ref{eq43})-(\ref{eq48}), to second-order terms on  $\epsilon$, that is
	\begin{equation}\label{eq44} \epsilon(1-\epsilon)=\frac{V}{3r_s^3}, \end{equation}
	where $V=r_s(0)^3-r(0)^3$ is initial volume of material.
substituting  value $\epsilon=1-r/r_s$ into (\ref{eq44}), we obtain quadratic equation for granule radius $r_s$, with the solution in the following form: :
	\begin{equation}\label{eq45} r_s=r\left(1+\frac{V}{3r^3}\right)  \end{equation}
Thus, within asymptotic approximation (\ref{eq42}), the connection is obtained between the pore and granule radii. Let us now proceed to the calculation of parameter $\alpha$, taking into account the condition $r \gg L$:
\begin{equation}\label{eq46} \alpha\approx\frac{r_s^2}{2L}\left(1+(1-\epsilon)^4-2(1-\epsilon)^2\right)^{1/2}=\frac{r_s^2\epsilon}{L}  \end{equation}
Hence, according to the definition (\ref{eq5}) one finds bispherical coordinates   $\eta_{1,2}$:
\begin{equation}\label{eq47} \quad \eta _1  = \textrm{arsinh}\left(\frac{r_s^2\epsilon}{rL}\right),\quad
\eta _2 = \textrm{arsinh}\left(\frac{r_s}{L}\epsilon\right) \end{equation}
 Because of the geometrical conditions, the inequality $\epsilon r_s/L\geq 1$ is valid. Thus, bispherical coordinates $\eta_{1,2}$ can be approximated for the case $\epsilon r_s/L \gg 1$ in the following form:
\begin{equation}\label{eq48}
\eta _1  \approx \ln \left(\frac{2r_s^2\epsilon}{rL}\right),\quad \eta _2  \approx \ln \left(\frac{2r_s\epsilon}{L}\right) \end{equation}

\begin{figure}
  \centering
  \includegraphics[ height=7 cm]{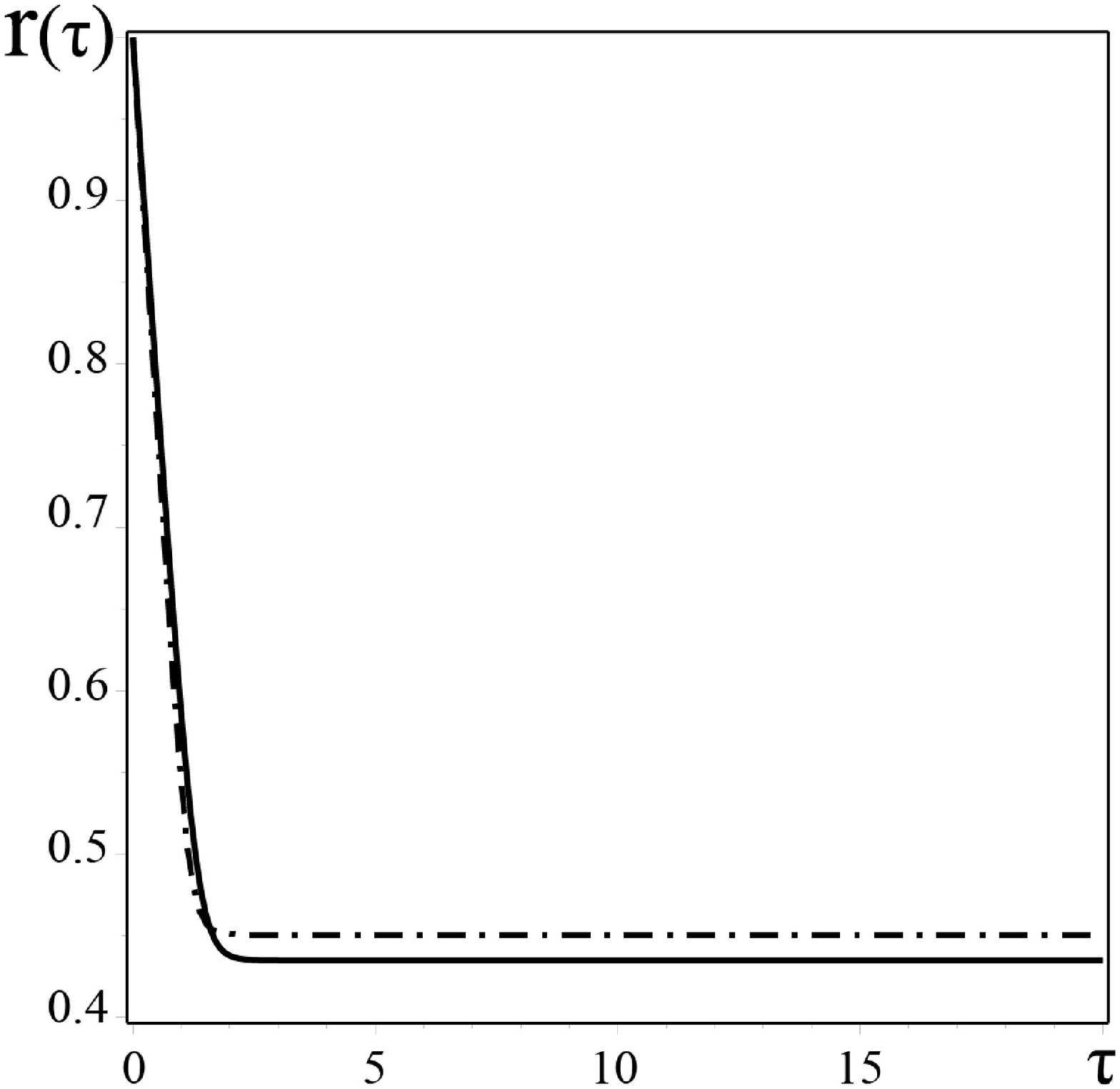}
	\includegraphics[ height=7 cm]{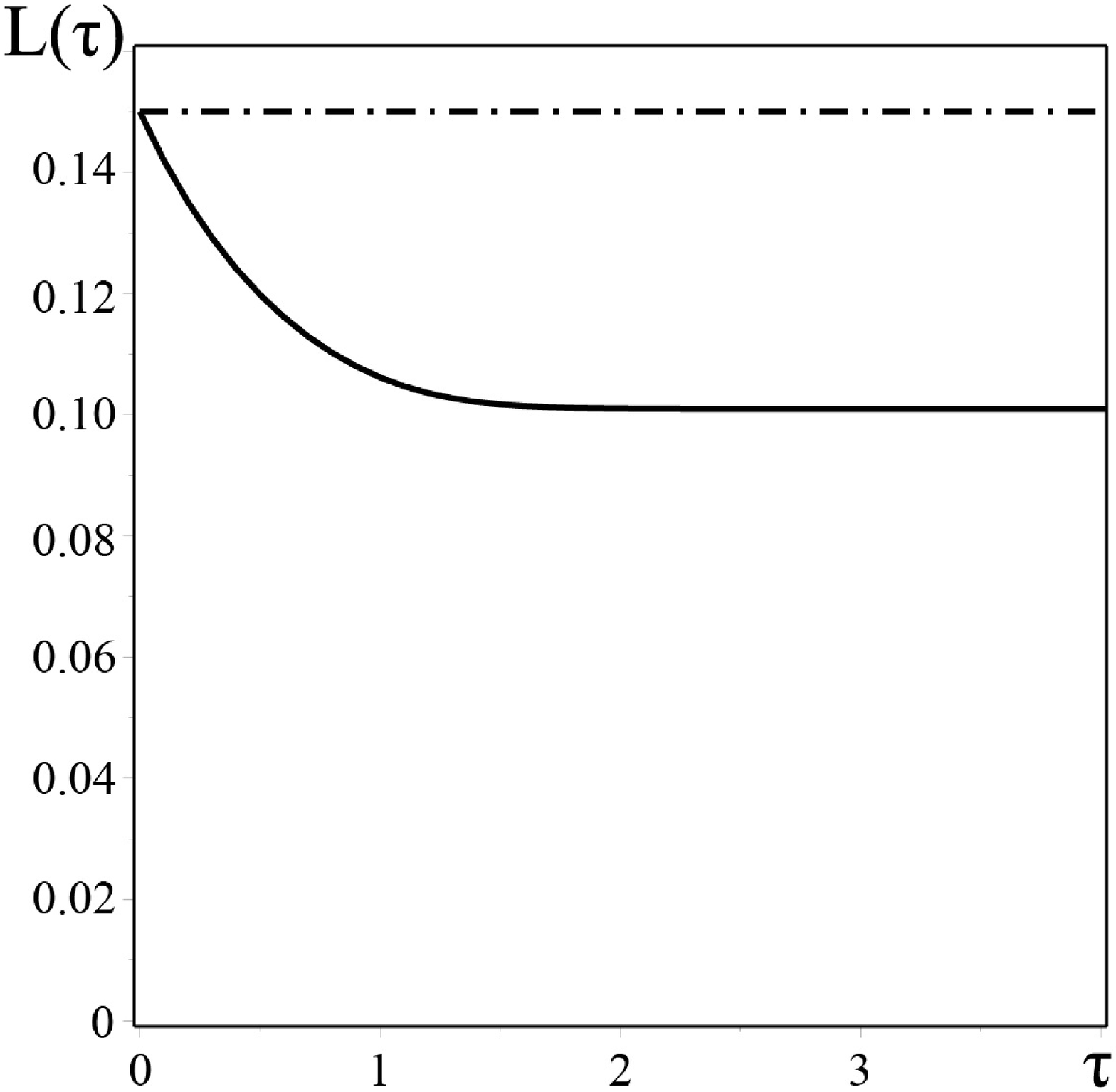}\\
\caption{On the left, solid line indicates the dependencies for large pore of pore radius $r$ on time $\tau$, obtained by numerical solution of equation set  (\ref{eq21}), dash-and-dot line relates to numerical solution of Eqs. (\ref{eq50})-(\ref{eq52}); on the right, dependence is shown of   distance time-change $L$ on time $\tau$. Solid line corresponds to the numerical solution of equation set (\ref{eq21}), while dash-and-dot line relates to solution of Eqs. (\ref{eq50})- (\ref{eq52}). All solutions are obtained for initial conditions $r|_{t=0}=1$, $r_s|_{t=0}=1.5$, $L|_{t=0}=0.15$, $A=10^{-1}$ and $B=0.025$.}
\label{fg9}
\end{figure}
Then, we find the difference  $\eta_1-\eta_2 =\ln \left(\frac{r_s}{r}\right) \approx \ln(1+\epsilon)\approx \epsilon$ and, correspondingly, make an estimate of series sums:
\[ \Phi_1 =\frac{1}{2\sinh(\eta_1+\epsilon)} \approx \frac{1}{2(\sinh\eta_1+\epsilon\cosh\eta_1) }\approx\frac{r}{2\alpha(1+\epsilon)}=\frac{r}{2\alpha}\left(1-\epsilon+\epsilon^2+\cdots\right),\]
\begin{equation}\label{eq49}
 \Phi_2 =\frac{1}{2\sinh(\eta_2+\epsilon)} \approx  \frac{1}{2(\sinh\eta_2+\epsilon\cosh\eta_2) }\approx \frac{r_s}{2\alpha(1+\epsilon)}=\frac{r_s}{2\alpha}\left(1-\epsilon+\epsilon^2+\cdots\right),
\end{equation}
\[ \widetilde{\Phi}_1 = \frac{\cosh(\eta_1+\epsilon)}{2\sinh^2(\eta_1+\epsilon)} \approx \frac{\cosh\eta_1+\epsilon\sinh\eta_1 }{2(\sinh\eta_1+\epsilon\cosh\eta_1)^2 } \approx \frac{r}{2\alpha(1+\epsilon)}= \frac{r}{2\alpha}\left(1-\epsilon+\epsilon^2+\cdots\right),\]
\[ \widetilde{\Phi}_2 =\frac{\cosh(\eta_2+\epsilon)}{2\sinh^2(\eta_2+\epsilon)} \approx \frac{\cosh\eta_2+\epsilon\sinh\eta_2 }{2(\sinh\eta_2+\epsilon\cosh\eta_2)^2 } \approx \frac{r_s}{2\alpha(1+\epsilon)}=\frac{r_s}{2\alpha}\left(1-\epsilon+\epsilon^2+\cdots\right). \]
 Substituting relations (\ref{eq45}), (\ref{eq46}) and (\ref{eq48})into equation set  (\ref{eq21}) we obtain evolution equations for a large pore with the accuracy up to second order term  $\epsilon^2$:
\begin{equation}\label{eq50} \frac{dL}{d\tau}=O(\epsilon^3) \end{equation}
\begin{equation}\label{eq51} \frac{dr}{d\tau}=-\frac{\exp\left(\frac{A}{r}-\frac{B}{r^3} \right)}{r}\cdot\left[\frac{1}{2}+\frac{1}{2}\cdot\left(\frac{2-\epsilon}{1-\epsilon}\right)(1-\epsilon+\epsilon^2)\right]+\frac{\exp\left(-\frac{A}{r\left(1+\frac{V}{3r^3}\right)}\right)}{r}\cdot\frac{1-\epsilon+\epsilon^2}{1-\epsilon} \end{equation}
It follows from Eq. (\ref{eq51}) that, within the considered asymptotic approximation, the change of distance $L(\tau)$ is quite small: $L(\tau) \approx L(0)$. Eq. (\ref{eq51}) does not depend on $L(\tau)$ , thus, substituting into it the value $\epsilon=1-r/r_s$ we find pore radius evolution equation:
\begin{equation}\label{eq52}	 \frac{dr}{d\tau}=-\frac{\exp\left(\frac{A}{r}-\frac{B}{r^3} \right)}{r}\cdot\left[1+\frac{1}{2}\cdot\frac{1}{1+\frac{V}{3r^3}} +\left(\frac{\frac{V}{3r^3}}{1+\frac{V}{3r^3}}\right)^2 \right]+$$
$$+\frac{\exp\left(-\frac{A}{r\left(1+\frac{V}{3r^3}\right)}\right)}{r}\cdot\left(1+\left(\frac{\frac{V}{3r^3}}{1+\frac{V}{3r^3}}\right)^2\right) \end{equation}
In Fig. \ref{fg9} dash-and-dot  line indicates numerical solution of approximate equations (\ref{eq50})-(\ref{eq52}) with initial conditions  $r|_{\tau =0}=1$, $r_s|_{\tau =0}=1.5$, $L|_{\tau=0}=0.15$, $A=10^{-1}$ and values of gas parameter $B=0.025$. It can be seen from Fig. \ref{fg9}, that numerical solutions of exact (\ref{eq21})  and  approximate (\ref{eq50})-(\ref{eq52}) well agree with each other at "low" gas pressure  inside the pore. Here, deviation $\Delta L(\tau)$  does not exceed accuracy order $\Delta L(\tau) \ll \epsilon^2$.

\subsection{Gas-filled pore in the center of spherical granule}

Here, we will consider the simplest limiting case $(l=0)$, when gas-filled pore is situated in the center of spherical granule . Geometrical inequality (\ref{eq22}) turns into evident one: $R<R_s$. It is convenient to consider this case in a spherical coordinate system. Boundary conditions for concentration remain the same and are determined by formulas (\ref{eq1}) and (\ref{eq2})correspondingly. Then equations, determining vacancy concentration  and boundary conditions, with account of the symmetry of the problem, take on a simple form:
\begin{equation}\label{eq53}	
\Delta_r c = 0, \quad c(r)|_{r=R}=c_R, \quad  c(r)|_{r=R_s}=c_{R_s},
\end{equation}
where $\Delta_r= \frac{1}{r^2}\frac{d}{d r}\left(r^2\frac{d}{d r}\right)$ is radial part of laplacian in spherical coordinates.
One can easily find the expression for vacancy concentration from Eq. (\ref{eq53}):
\begin{equation}\label{eq54}	c(r)=-\frac{C_1}{r}+C_2, \end{equation}
where $C_{1,2}$ are arbitrary constants, that are determined by boundary conditions. Vacancy flux $\vec j$ is determined by the first  Fick's law  (\ref{eq11}), while vacancy fluxes per unit surface of the pore or the granule equal, correspondingly, to
 \begin{equation}\label{eq55}
\vec n \cdot \vec j|_{r=R}=\frac{D}{\omega}\frac{\partial c}{\partial r}|_{r=R},\quad \vec n \cdot \vec j|_{r=R_s}=\frac{D}{\omega}\frac{\partial c}{\partial r}|_{r=R_s}
\end{equation}
Substituting these expressions into the equation for the change of the volume of the pore and the granule
\[\dot{R}=-\frac{\omega}{4\pi R^2} \oint\vec{n}\vec{j}|_{r=R} dS, \; \dot{R_s}=-\frac{\omega}{4\pi R_s^2}\oint \vec{n}\vec{j}|_{r=R_s} dS   \]
we find the equation  for time change of the radius of the pore and the granule:
\begin{equation}\label{eq56}
\left\{
\begin{aligned}
 \dot{R}=-\frac{D}{R}\cdot\frac{(c_{R_s}-c_R)R_s}{R_s-R} \\
  \dot{R_s}=-\frac{D}{R_s}\cdot\frac{(c_{R_s}-c_R)R}{R_s-R} \\
\end{aligned}
\right.
\end{equation}
It can be checked easily, that, from the evolution equation  (\ref{eq56}) for the gas-filled pore in the granule center, the conservation law follows:
\[R_s(t)^2\dot{R}_s(t)-R(t)^2\dot{R}(t)=0.\]
Thus,  granule radius  is connected with pore pore volume by the simple relation:
\begin{equation}\label{eq57}
  R_s(t) =\sqrt[3]{V+R(t)^3},
\end{equation}
where $V=R_s(0)^3 -R(0)^3$ is initial volume of granule material. Critical radius of the pore is determined from the first equation of the set (\ref{eq56}), assuming  $\dot{R}=0$ and using formulas (\ref{eq1})-(\ref{eq2}) and ideal gas equation:
\begin{equation}\label{eq58}
R_{cr}=\frac{1}{2}\sqrt{\frac{3N_g kT}{2\pi\gamma}}\end{equation}
As it can be seen from (\ref{eq58}), for vacancy pore with $(N_g=0)$, threre exists no stationary radius $(R_{cr}=0)$, that agrees with the conclusions of the work \cite{21s}. Moreover, we see that stationary radius of gas-filled pore grows with an increase of concentration $N_g$ and temperature $T$ on the gas. Speaking generally, the value of stationary radius retains also in a more complicated case of an  arbitrarily situated pore.

\section{Conclusions}

Let us finally discuss general regularities of the behaviour of a gas-filled pore  inside a spherical granule in hydrodynamical approximation. First of all, in the limiting case of the absence of gas-diffusion in the matrix, there exists stationary pore radius, that is ultimately reached by the pore.  This stationary radius  is determined by the quantity of gas inside the pore as well as by granule temperature. Thus, depending on the relation between stationary radius value and initial pore radius,  pore size can either increase or decrase  with time. In particular case of pore radius coincidence of initial radius with the stationary one, pore size does not change. In general case, gas-filled pore shifts towards granule center if its size is diminishing down to stationary value or away from granule center if its size is increasing up to stationary value. It should be noted, that such shift is small since pore motion stops as soon as pore radius reaches its stationary value. The particular case of coaxial  position of the pore in the granule yields simple equations for pore and granule size change, that are in a good agreement with the more complicated case  of arbitrary position of the pore in the spherical granule.

\section*{Appendix}

\subsection*{1. Auxillary relations.}
\begin{equation}\label{EQ1}
\int_{-1}^1\frac{P_k(t)dt}{\sqrt{\cosh\eta-t}} = \frac{\sqrt{2}\cdot
e^{-(k+1/2)\eta}}{k+1/2}\,.
\end{equation}
 Differentiating relation (\ref{EQ1}) with respect to the parameter $\eta$, we subsequently find
\begin{equation}\label{EQ2}
\int_{-1}^1\frac{P_k(t)dt}{(\cosh\eta-t)^{3/2}} =
\frac{2\sqrt{2}\cdot e^{-(k+1/2)\eta}}{\sinh\eta}\,,
\end{equation}
\begin{equation}\label{EQ3}
\int_{-1}^1\frac{P_k(t)dt}{(\cosh\eta-t)^{5/2}} =
\frac{4\sqrt{2}\cdot
e^{-(k+1/2)\eta}(\cosh\eta+(k+1/2)\sinh\eta)}{3\cdot\sinh^3\eta}\,.
\end{equation}
\subsection*{2. Calculation of pore radius change .}

\begin{equation}\label{EQ4} \dot{R}=-\frac{\omega}{4\pi
R^2}\oint\vec{n}\vec{j}dS\,,
\end{equation}
 where
\begin{equation}\label{EQ5}
\vec{n}\vec{j}|_{\eta=\eta_1}=\frac {D}{\omega} \cdot
\frac{\cosh\eta_1-\cos\xi}{a}\frac{\partial
c}{\partial\eta}|_{\eta=\eta_1}\,,
\end{equation}
 \begin{equation}\label{EQ6}
 dS = \frac{a^2\cdot \sin\xi
d\xi d\varphi}{(\cosh\eta_1-\cos\xi)^2}\,,
\end{equation}
 After application of Fubini's theorem, with account of the independence of
 $\xi$ and $\varphi$ the expression for $\dot{R}$ takes on a form
 \begin{equation}\label{EQ7}
  \dot{R}=-\frac{a\cdot D}{2\cdot
R^2}\int_0^{\pi}\frac{\partial
c}{\partial\eta}|_{\eta=\eta_1}\frac{\sin\xi d\xi
}{\cosh\eta_1-\cos\xi}\,,
\end{equation}
 Substituting
$$\frac{\partial
c}{\partial\eta}|_{\eta=\eta_1} = \sqrt{2}\left(
\frac{c_{R}\cdot\sinh\eta_1}{\sqrt{{\cosh\eta_1-\cos\xi}}}\cdot\sum_{k=0}^{\infty}
P_k(\cos\xi)\exp(-\eta_1(k+1/2))+
\sqrt{{\cosh\eta_1-\cos\xi}}\times\right.$$
$$\left.\times\sum_{k=0}^\infty\frac{(k+1/2)\cdot
P_k(\cos\xi)}{\sinh(k+1/2)(\eta_1-\eta_2)}\left[c_{R}\cdot\cosh(k+1/2)(\eta_1-\eta_2)e^{-\eta_1(k+1/2)}
-c_{R_s}\cdot e^{-\eta_2(k+1/2)}\right] \right)$$
into the expression for the speed of pore radius change, and exchanging integration and summation signs on the strength of convergence of corresponding sums and integrals, after substituting $\cos\xi=t$, we obtain
$$ \dot{R}=-\frac{a\cdot D\sqrt{2}}{2\cdot
R^2}\left[\frac{c_{R}\cdot\sinh\eta_1}{2}\sum_{k=0}^\infty
e^{-\eta_1(k+1)}\int_{-1}^1\frac{P_k(t)dt}{(\cosh\eta_1-t)^{3/2}}+\right.$$
$$\left. +\sum_{k=0}^\infty\frac{(k+1/2)}{\sinh(k+1/2)(\eta_1-\eta_2)}\left[c_{R}\cdot\cosh(k+1/2)
(\eta_1-\eta_2)e^{-\eta_1(k+1/2)} -\right.\right. $$
$$\left.\left.- c_{R_s}\cdot
e^{-\eta_2(k+1/2)}\right]\cdot
\int_{-1}^1\frac{P_k(t)dt}{\sqrt{\cosh\eta_1-t}}\right] \,.$$
Using values of integrals (\ref{EQ1}) and (\ref{EQ2}), we can reformulate this expression
$$ \dot{R}=-\frac{a\cdot D\sqrt{2}}{2\cdot
R^2}\left[\frac{c_{R}\cdot\sinh\eta_1}{2}\sum_{k=0}^\infty
e^{-\eta_1(k+1/2)}\cdot \frac{2\sqrt{2}\cdot
e^{-\eta_1(k+1/2)}}{\sinh\eta_1} +\right.$$
$$\left. +\sum_{k=0}^\infty\frac{(k+1/2)}{\sinh(k+1/2)(\eta_1-\eta_2)}\left[c_{R}\cdot\cosh(k+1/2)
(\eta_1-\eta_2)e^{-\eta_1(k+1/2)} -\right.\right. $$
$$\left.\left.- c_{R_s}\cdot
e^{-\eta_2(k+1/2)}\right]\cdot \frac{\sqrt{2}\cdot
e^{-(k+1/2)\eta_1}}{k+1/2} \right]= -\frac{a\cdot D}{R^2}\left[
c_{R}\cdot\sum_{k=0}^\infty e^{-\eta_1(2k+1)} +\right.$$
$$\left. +\sum_{k=0}^\infty \frac{c_{R}\cdot\cosh(k+1/2)
(\eta_1-\eta_2)e^{-\eta_1(k+1/2)} + c_{R_s}\cdot e^{-\eta_2(k+1/2)}}
{\sinh(k+1/2)(\eta_1-\eta_2)}\cdot e^{-\eta_1(k+1/2)}\right] =
-\frac{a\cdot D}{R^2}\times$$
$$\times\left[\frac{c_{1}}{2\cdot\sinh\eta_1} +\sum_{k=0}^\infty
\frac{c_{R}\cdot\cosh(k+1/2) (\eta_1-\eta_2)e^{-\eta_1(k+1/2)} -
c_{R_s}\cdot e^{-\eta_2(k+1/2)}} {\sinh(k+1/2)(\eta_1-\eta_2)}\cdot
e^{-\eta_1(k+1/2)}\right]\,.$$
 Substituting  $a=R\cdot \sinh\eta_1$ and transforming summing terms,\\
 we ultimately obtain:
\begin{equation}\label{EQ8} \dot{R}=-\frac{D}{R}\left[\frac{c_{R}}{2}
+\sinh\eta_1\cdot\sum_{k=0}^\infty
\frac{c_{R}\cdot(e^{-(2k+1)\eta_1}+e^{-(2k+1)\eta_2}) -2\cdot
c_{R_s}\cdot e^{-(2k+1)\eta_2}}
{e^{(2k+1)(\eta_1-\eta_2)}-1}\right]\,.
\end{equation}

 \subsection*{3. Calculation of the speed of pore motion.}

$$ \vec{v}=\vec{e_z}\cdot\frac{3\cdot D\cdot
a}{2\cdot R^2}\int_0^\pi \frac{\partial
c}{\partial\eta}|_{\eta=\eta_1}\frac{\cosh\eta_1\cdot
\cos\xi-1}{(\cosh\eta_1-\cos\xi)^2}\cdot\sin\xi d\xi = $$
$$=\vec{e_z}\cdot\frac{3\cdot D\cdot
a}{2\cdot R^2}\int_0^\pi \frac{\partial
c}{\partial\eta}|_{\eta=\eta_1}\left(-\frac{\cosh\eta_1}{\cosh\eta_1-\cos\xi}+
\frac{\sinh^2\eta_1}{(\cosh\eta_1-\cos\xi)^2}\right) \cdot\sin\xi
d\xi =$$
$$=\vec{e_z}\cdot\frac{3\sqrt{2}\cdot D\cdot
a}{2\cdot R^2}
\int_0^\pi\left(-\frac{\cosh\eta_1}{\cosh\eta_1-\cos\xi}+
\frac{\sinh^2\eta_1}{(\cosh\eta_1-\cos\xi)^2}\right) \cdot\sin\xi
d\xi\times$$
$$\times
\left[ \frac{c_R\cdot\sinh\eta_1}{2\cdot\sqrt{\cosh\eta_1-\cos\xi}}\cdot\sum_{k=0}^\infty P_k(\cos\xi)
e^{-\eta_1(k+1/2)} + \sqrt{\cosh\eta_1-\cos\xi}\times\right.$$

$$\left.\times\left( \sum_{k=0}^\infty\frac{(k+1/2)
P_k(\cos\xi)(c_R\cdot e^{-\eta_1(k+1/2)}\cosh(k+1/2)(\eta_1-\eta_2) -c_{R_s}\cdot
e^{-\eta_2(k+1/2)})}{\sinh(k+1/2)(\eta_1-\eta_2)}
 \right)\right]\,.$$
The substitution $t=\cos\xi$ and exchange of summation and integration signs yield the expression
$$\vec{v}=\vec{e_z}\cdot\frac{3\sqrt{2}\cdot D\cdot
a}{2\cdot R^2}
\sum_{k=0}^\infty\left[\int_{-1}^1\frac{P_k(t)dt}{(\cosh\eta_1-t)^{5/2}}\cdot
\frac{c_R\cdot e^{-\eta_1(k+1/2)}\sinh^3\eta_1}{2}
\right.+$$
$$+\int_{-1}^1\frac{P_k(t)dt}{(\cosh\eta_1-t)^{3/2}}\times\left(\frac{-c_R\cdot e^{-\eta_1(k+1/2)}
\cosh\eta_1\sinh\eta_1}{2}\right.+\sinh^2\eta_1\times$$
$$\left.
\times\frac{(k+1/2)
(c_R\cdot e^{-\eta_1(k+1/2)}\cosh(k+1/2)(\eta_1-\eta_2) -c_{R_s}\cdot
e^{-\eta_2(k+1/2)})}{\sinh(k+1/2)(\eta_1-\eta_2)}
 \right)+\int_{-1}^1\frac{P_k(t)dt}{\sqrt{\cosh\eta_1-t}}\times$$
$$ \left.\times\left(
-\cosh\eta_1\cdot\frac{(k+1/2)
(c_R\cdot e^{-\eta_1(k+1/2)}\cosh(k+1/2)(\eta_1-\eta_2) -c_{R_s}\cdot
e^{-\eta_2(k+1/2)})}{\sinh(k+1/2)(\eta_1-\eta_2)}
 \right)\right]\,.$$
Now le us substitute values of corresponding integrals \\ into the obtained expression and reform the result
$$\vec{v}=\vec{e_z}\cdot\frac{3\sqrt{2}\cdot D\cdot
a}{2\cdot R^2} \sum_{k=0}^\infty\left[\frac{4\sqrt{2}\cdot
e^{-(k+1/2)\eta_1}(\cosh\eta_1+(k+1/2)\sinh\eta_1)}{3\cdot\sinh^3\eta_1}\times
\right. $$
 $$\times\frac{c_R\cdot e^{-\eta_1(k+1/2)}\sinh^3\eta_1}{2}+\frac{2\sqrt{2}\cdot
e^{-(k+1/2)\eta_1}}{\sinh\eta_1}\times
\left(\frac{-c_R\cdot e^{-\eta_1(k+1/2)}\cosh\eta_1}{2}
+\sinh^2\eta_1\times\right.$$
$$\left. \times\frac{(k+1/2)
(c_R\cdot e^{-\eta_1(k+1/2)}\cosh(k+1/2)(\eta_1-\eta_2) -c_{R_s}\cdot
e^{-\eta_2(k+1/2)})}{\sinh(k+1/2)(\eta_1-\eta_2)}
 \right)+\frac{\sqrt{2}\cdot e^{-(k+1/2)\eta_1}}{k+1/2}\times$$
$$\left. \times\left(
-\cosh\eta_1\cdot\frac{(k+1/2)
(c_R \cdot e^{-\eta_1(k+1/2)}\cosh(k+1/2)(\eta_1-\eta_2) -c_{R_s}\cdot
e^{-\eta_2(k+1/2)})}{\sinh(k+1/2)(\eta_1-\eta_2)}
 \right)\right]=$$
\begin{equation}\label{eq53s}=\vec{e_z}\cdot\frac{3 D
a}{ R^2}
\sum_{k=0}^\infty\frac{((2k+1)\sinh\eta_1-\cosh\eta_1)\left(c_R\cdot(e^{-(2k+1)\eta_1}+e^{-(2k+1)\eta_2})-2c_{R_s}\cdot
e^{-(2k+1)\eta_2}\right)}{e^{(2k+1)(\eta_1-\eta_2)}-1}\,.
\end{equation}


\begin{thebibliography}{23}



\bibitem{1s} V.V.  Slezov.  Coalescence of the system dislocation loops and pores under irradiation, Solid State Physics, 1967, Vol. 9, No. 12, p.3448.
\bibitem{2s}  V.V. Slezov, V.B.  Shikin.  On the coarsening of pores in solids with sources of gas, Euro nuclears. 1965, vol.2,  No. 3, pp.127-131.
\bibitem{3s}  Z.K.  Saralidze,  V.V. Slezov. Coalescence of dislocation loops in the nonstationary regime, Solid State Physics, 1965, Vol. 7, No. 3, p.1605.
\bibitem{4s} V.V. Slezov and V.V. Sagalovich.  Diffusive decomposition of solid solutions, Sov. Phys. Usp.30 (1987), pp. 23-45.
\bibitem{5s} V.V. Slezov, O.A. Osmayev, R.V. Shapovalov. An evolution of voids ensamble in an irradiated material, Problems  of Atomic Science and Technology, 2008, No.2,  Series: Physics of Radiation Effects and Radiation Materials Science (92), pp. 3-9.
\bibitem{6s} V.V. Slezov, O.A. Osmayev, R.V. Shapovalov. Bulbs moving  in a  material with gas atom sources, Problems  of Atomic Science and Technology, 2005. No. 3, Series: Physics of Radiation Effects and Radiation Materials Science (86), pp. 38-42.
\bibitem{7s} P.G. Cheremskoy, V.V. Slyozov, and V.I. Betehin, Pores in Solid Matter, Energoatomizdat, Moscow, 1990 (in Russian).
\bibitem{8s} Y. Yin, R. M. Rioux, C. K. Erdonmez, S. Hughes, G.A., A.P, Formation of hollow nanocrystals through the nanoscale Kirkendall effect. Science, 304 (2004), pp.711-714.
\bibitem{9s} C. M. Wang, D. R. Baer, L. E. Thomaset. al., Void formation during early stages of passivation: Initial oxidation of iron nanoparticles at room temperature, J. Appl. Phys., 98 (2005),  pp. 94308-94308.
\bibitem{10s}  A. Cabot, V. F. Puntes, E. Shevchenko et al., Vacancy Coalescence during Oxidation of Iron Nanoparticles, J. Am. Chem. Soc. 129, No. 34 (2007): 10358-10360.
\bibitem{11s} H. J. Fan, M. Knez, R. Scholz et al., Influence of Surface Diffusion on the Formation of Hollow Nanostructures Induced by the Kirkendall Effect: The Basic Concept, Nano Lett 7, No. 4 (2007), pp. 993-997.
\bibitem{12s} R. Nakamura, J. G. Lee, D. Tokozakura et al., Formation of hollow ZnO  through low temperature oxidation of Zn particles,  Mater. Lett. 61 (2007), pp. 1060-1063.
\bibitem{13s} R. Nakamura, D. Tokozakura, H. Nakajima et al., Hollow oxide formation byoxidation of  Aland Cu nanoparticles, J. Appl. Phys. 101 (2007), pp. 07430.
\bibitem{14s} D. Tokozakura, R. Nakamura, H. Nakajima et al., Transmission electron microscopy observation of oxide layer growth on Cu nanoparticles and formation process of hollow oxide particles,  Mater. Res., 22, No. 10 (2007), pp.2930-2935.
\bibitem{15s} R. Nakamura, J. G. Lee, H. Morix, and H. Nakajima, Oxidation behavior of Ni nanoparticles and formation process of hollow NiO, Philos. Mag. 88, No. 2 (2008), pp. 257-264.
\bibitem{16s}  R. Nakamura, D. Tokozakura, J.-G. Lee et al., Shrinking of hollow Cu2O and NiO nanoparticles at high temperatures, Acta Mater. 56 (2008), pp. 5276 - 5284.
\bibitem{17s}  R. Nakamura, G. Matsubayashi, H. Tsuchiya et al., Formation of oxide nanotubes via oxidation of Fe, Cu and Ni nanowires and their structural stability: difference in formation and shrinkage behavior of interior pores, Acta Mater. 57 (2009), pp. 5046-5052.
\bibitem{18s} V.V. Kulish. Composite Nanostructures  with Metal Components, J. Nano- Electron. Phys., 2011, V.3, No. 3, pp. 114-126.
\bibitem{19s} An Kwangjin, Hyeon Taeghwan. Synthesis and biomedical applications of hollow nanostructures. Nano Today (2009) 4, 359?73.
\bibitem{20s}  T.V. Zaporozhets, A.M. Gusak, and O.N. Podolyan, Evolution of Pores in Nanoshells:Competition of Direct and Reverse Effects of Kirkendall, Effects of Frenkel and Gibbs-Tomson (Phenomenological Description and Computer Simulation), Usp. Fiz. Met. 13 (2012), pp.1-70.
\bibitem{21s} V.V. Yanovsky , M.I. Kopp  and  M. A. Ratner. Evolution of vacancy pores in bounded particles. ArXiv:1809.06565v1[cond-mat.mes-hall]  (2018)
\bibitem{22s} Ja.E. Geguzin and M.A. Krivoglaz. Motion of Macroscopic Inclusions in Solid Matter, Metallurgy, Moscow, 1971 (in Russian).
\bibitem{23s} G. Arfken, Mathematical methods for physicists, Acad. Press, New York and London, 1970.



\end{thebibliography}
\end{document}